\documentclass[12pt,a4wide]{article}
\usepackage{amsmath,amssymb,amsfonts}
\newcommand{\eqa}{\begin{eqnarray}}
\newcommand{\ena}{\end{eqnarray}}

\newcommand{\topstar}[1]{\setlength{\unitlength}{1mm}
\begin{picture}(2,0)(-1,-1.4)
   \put(0,0){\makebox(0,0){$#1$}}
   \put(0,2.4){\makebox(0,0){\mbox{\tiny$\star$}}}
\end{picture}}
\begin{document}
\topmargin 0pt
\oddsidemargin 5mm
 \renewcommand{\thefootnote}{\fnsymbol{footnote}}
\newpage                                        
\setcounter{page}{1}
\begin{center}
{\large {\bf Spherically Symmetric, Metrically Static, Isolated
Systems in Quasi-Metric Gravity}}
\end{center}
\begin{center}
Dag {\O}stvang \\
{\em Department of Physics, Norwegian University of Science and Technology 
(NTNU) \\
N-7491 Trondheim, Norway}
\end{center}
\begin{abstract}
Working within the quasi-metric framework (QMF) described elsewhere, we 
examine the gravitational field exterior respectively interior to a
spherically symmetric, isolated body made of perfect fluid. By construction,
the system is ``metrically static'', meaning that its associated gravitational 
field is static except for the effects of the global cosmic expansion on the 
spatial geometry. To ensure that the global cosmic expansion will not induce 
instabilities in the fluid source and thus violating the metrically static 
condition, the equation of state of the fluid is required to take a particular 
form (fulfilled for, e.g., an ideal gas). 

We set up dynamical equations for the gravitational field and give an exact 
solution for the exterior part. Furthermore, we find equations of motion to be
applied to inertial test particles moving in the exterior gravitational 
field. The metrically static condition implies that the radius of the source 
increases and that distances between circular orbits increase according to the 
Hubble law, but such that circle orbit velocities are unaffected. This means 
that the dynamically measured mass of the source increases linearly with 
cosmic scale. We show that, if this model of an expanding gravitational field 
is taken to represent the gravitational field of the solar system, this has no 
serious consequences for observational aspects of planetary motion. On the 
contrary, some observational facts of the Earth-Moon system are naturally 
explained within the QMF. Finally, the QMF predicts different secular increases 
for two different gravitational coupling parameters. But such secular changes 
are neither present in the Newtonian limit of the quasi-metric equations of 
motion nor in the Newtonian limit of the quasi-metric field equations valid 
inside metrically static sources. Thus standard interpretations of space 
experiments, testing the secular variation of the gravitational ``constant'', 
are explicitly theory-dependent and do not apply to the QMF.
\end{abstract}
\section{Introduction}
The idea that the cosmic expansion may possibly be relevant for local systems 
came up many years ago; see, e.g., [1] and references therein. More recently,
there has been renewed interest in this idea; in part because it has become 
clear that there is no compelling observational evidence showing the expected 
deviations from the global Hubble law for galaxies in the vicinity of or even
within the local group of galaxies. See, e.g., [2] and references therein.

Even if there are in principle no direct observations ruling out the relevance
of the global Hubble law on local scales, the generally accepted view is that 
for all practical purposes, local systems may be treated as decoupled from the
cosmic expansion. This view reflects predictions coming from the standard 
framework of metric gravity. That is, it is well-known that metric theory 
predicts that realistic local systems are hardly affected at all by the 
cosmological expansion (its effect should at best be totally negligible, see,
e.g., [1] and references therein). The reason for this prediction is basically 
that in metric theory, the cosmological expansion must be modelled within a 
mathematical framework where space-time is postulated to be a pseudo-Riemannian 
manifold. However, when analyzing the influence of the cosmological expansion 
on local systems, there should be no reason to expect that predictions made 
within the metric framework should continue to hold in a theory where the 
structure of space-time is non-metric.

Recently, a review of a new type of non-metric space-time framework, the 
so-called quasi-metric framework (QMF), was presented in [3]. Also presented 
was an alternative relativistic theory of gravity formulated within this 
framework. (A more detailed presentation can be found in [4].) This theory 
correctly predicts the results of the ``classical'' solar system tests in the 
so-called ``metric approximation'' case where an asymptotically Minkowski 
background is invoked and the non-metric features of the theory (and thus the 
cosmological expansion) can be neglected. However, for reasons explained in 
[3], the theory is based on a ${\bf S}^3{\times}{\bf R}$-background rather 
than a Minkowski background as the global basic (``prior'') geometry of the 
Universe. As long as the cosmological expansion is neglected, the choice of 
cosmic background geometry does not matter for the predictions of said solar 
system tests, though.

Moreover, since it represents the non-metric sector of the theory, in 
quasi-metric theory the nature of the cosmological expansion is described as 
fundamentally different from its counterpart in metric theory. One consequence 
of the quasi-metric description of the cosmological expansion is that the 
expansion applies to all systems where gravitational dynamics dominates 
(hereafter called ``gravitational systems''), regardless of scale. That is, in 
quasi-metric gravity the mathematical modelling of the Hubble expansion and 
thus its physical interpretation are different from their counterparts in 
metric theory, and as a consequence, the Hubble expansion is predicted to 
influence local, gravitationally bound systems sufficiently that its effects 
should be observable in experiments. On the other hand, since 
quantum-mechanical states should be unaffected by the expansion, quasi-metric 
theory allows that the global cosmic expansion does not apply to 
quantum-mechanical systems bound by non-gravitational forces where 
gravitational interactions are negligible (hereafter called 
``atomic systems'') [5].

To find more exactly how the cosmological expansion affects local 
gravitational systems according to the quasi-metric theory, one must first 
calculate the spherically symmetric gravitational field with the 
${\bf S}^3{\times}{\bf R}$-background, both interior and exterior to the 
source. Then one should use the quasi-metric equations of motion (with their 
non-metric terms included) to calculate how test particles move in the exterior
gravitational field. We show in section 3.3 of this paper that the quasi-metric
theory predicts that the exterior gravitational field should expand according 
to the Hubble law. This also applies to the interior gravitational field if 
potential instabilities induced by the global cosmic expansion can be neglected
(see section 3.2). In particular, for a source made of ideal gas, the cosmic 
expansion induces no instabilities so the radius of a body made of ideal gas is
predicted to expand. This result may support an interpretation of geological 
data indicating that the Earth is expanding according to the Hubble law, see 
reference [6] and references cited therein. (It is difficult to measure such a 
small expansion rate directly due to the existence of larger local 
displacements of the Earth's surface.) Besides, an expanding Earth should 
cause changes in its spin rate; we show in section 4.2 of this paper that the 
secular spin-down of the Earth as inferred from historical astronomical 
observations may in fact be of cosmological origin and only about half of the 
currently accepted value. Quasi-metric theory also predicts a cosmological 
origin of and different values for the recession of the Moon and its mean 
acceleration, other than those inferred from lunar laser ranging (LLR) 
experiments using standard theory. However, these differences are due to 
model-dependence since the LLR data yields that the recession of the Moon 
follows Hubble's law when analyzed within the QMF, and the quasi-metric 
predictions are consistent with a modern lunar ephemeris.

Finally, it is shown that the predicted cosmic expansion of the solar system's 
gravitational field does not lead to easily detected perturbations in the 
observed motion of the planets. However, some less easily detected effects 
should be measurable; in fact a newly discovered secular increase of the 
astronomical unit may be explained by cosmic expansion. Also active mass is
predicted to show a secular increase; in section 4.3 we argue that the 
predicted value of this increase is not in conflict with current test 
experiments. All these results are very different from their counterparts in 
metric theory. On the other hand, predictions of and data analyses based
on metric theory are the reasons why it is generally believed that observations
confirm that the solar system is decoupled from the cosmic expansion when it 
is in fact the other way around.
\section{Quasi-metric relativity in brief}
\subsection{General formulae}
In this section we summarize the main features of the QMF and a quasi-metric 
theory of gravity. A considerably more extensive discussion can be found in 
[3] or [4].

The mathematical foundation of the QMF can be described by first considering a
5-dimensional product manifold ${\cal M}{\times}{\bf R}_1$, where
${\cal M}={\cal S}{\times}{\bf R}_2$ is a (globally hyperbolic) Lorentzian
space-time manifold, ${\bf R}_1$ and ${\bf R}_2$ are two copies of the real
line and ${\cal S}$ is a compact Riemannian 3-dimensional manifold (without
boundaries). Then {\em the global time function t} representing the extra
(degenerate) time dimension ${\bf R}_1$ is introduced as a coordinate on 
${\bf R}_1$. Moreover, for $t$ given it is convenient to use a coordinate 
system ${\{}x^{\mu}{\}}$ (${\mu}$ taking integer values in the range $0-3$) 
where the ordinary time coordinate $x^0$ on ${\cal M}$ scales like $ct$; this 
ensures that $x^0$ is in some sense a mirror of $t$ and thus a ``preferred'' 
global time coordinate. A coordinate system with a global time coordinate of 
this type we call a {\em global time coordinate system} (GTCS). Hence, 
expressed in a GTCS ${\{}x^{\mu}{\}}$, $x^0$ is interpreted as a global 
coordinate on ${\bf R}_2$ and ${\{}x^j{\}}$ ($j$ taking integer values in the 
range $1-3$) as spatial coordinates on ${\cal S}$. The class of GTCSs is a set 
of preferred coordinate systems inasmuch as the equations of quasi-metric 
relativity take special forms when expressed in a GTCS. Note that there exist 
infinitely many GTCSs. 

The 4-dimensional quasi-metric space-time manifold ${\cal N}$ can now be 
defined by slicing the sub-manifold $x^0=ct$ (using a GTCS) out of the initial 
5-dimensional space-time manifold. Furthermore, $\cal N$ is equipped with two 
families of Lorentzian space-time metric tensor fields ${\bf {\bar g}}_t$ and 
${\bf g}_t$. The ``dynamical'' metric family ${\bf {\bar g}}_t$ represents a 
solution of field equations, and from ${\bf {\bar g}}_t$ one can construct the 
``physical'' metric family ${\bf g}_t$ which is used when comparing predictions
to observations (involving the equations of motion). It is convenient to think 
of the metric families as single degenerate metrics on (a subset of) 
${\cal M}{\times}{\bf R}_1$, where the degeneracy manifests itself via the 
conditions ${\bf {\bar g}}_t({\frac{\partial}{{\partial}t}},{\cdot}){\equiv}0$ 
and ${\bf g}_t({\frac{\partial}{{\partial}t}},{\cdot}){\equiv}0$. Finally, 
note that ${\cal N}$ differs from a Lorentzian manifold and that this becomes 
evident only when it is equipped with an affine connection (see below).

From the above description we see that within the QMF, the canonical 
description of space-time is taken as fundamental. That is, quasi-metric 
space-time is constructed as consisting of two mutually orthogonal foliations:
on the one hand space-time can be sliced up globally into a family of 
3-dimensional space-like hypersurfaces (called the fundamental hypersurfaces 
(FHSs)) by the global time function $t$, on the other hand space-time can be 
foliated into a family of time-like curves everywhere orthogonal to the FHSs. 
These curves represent the world lines of a family of hypothetical observers 
called the fundamental observers (FOs). There exists a unique relationship 
between $t$ and the proper time as measured by any FO.

Now one characteristic property of quasi-metric theory is that it postulates
the existence of systematic scale changes between gravitational and atomic 
systems (and the main role of $t$ is to describe the global aspects of such
changes). This means that gravitational quantities are postulated to exhibit 
an extra variation when measured in atomic units (and {\em vice versa}). One 
may think of this as if fixed operationally defined atomic units vary formally 
in space-time. Moreover, since $c$ and Planck's constant ${\hbar}$ by 
definition are not formally variable, the formal variation of time units is 
equal to that of length units and inverse to that of mass units. We now 
postulate that this formal variation of atomic length (or time) units can be 
defined from a particular geometric feature of the FHSs in 
$({\cal N},{\bf {\bar g}}_t)$. That is, said formal variation is defined in 
terms of the variation of the spatial scale factor ${\bar F}_t$ of the FHSs 
being a distinctive geometric feature of ${\bf {\bar g}}_t$. Thus by 
definition, measured in atomic units, the formal variability of gravitational 
quantities with the dimension of time or length goes as ${\bar F}_t$, whereas 
the the formal variability of gravitational quantities with the dimension of 
mass goes as ${\bar F}_t^{-1}$ (gravitational quantities with the dimension of 
charge have no formal variability).

To determine the form of ${\bar F}_t$, we require that no extra arbitrary scale 
or parameter should be introduced (i.e., no characteristic scale should be 
associated with ${\bar F}_t$). This yields the (rather unique) choice
${\bar F}_t{\equiv}c{\bar N}_tt$, where ${\bar N}_t$ is the lapse function
field family of the FOs in $({\cal N},{\bf {\bar g}}_t)$. With ${\bar F}_t$
given, together with the requirement that the FHSs should be compact and have
a trivial topology, it is now straightforward to set up the general form of 
${\bf {\bar g}}_t$. Thus it can be argued [3, 4] that, expressed in component
notation in a suitable GTCS, the most general form allowed for the family 
${\bf {\bar g}}_t$ may be represented by the family of line elements (we use 
the metric signature $(-+++)$ and Einstein's summation convention throughout)
\eqa
{\overline {ds}}_t^2={\bar N}_t^2{\Big \{ }
[{\bar N}_{(t)}^k{\bar N}_{(t)}^s{\tilde h}_{(t)ks}-1](dx^0)^2+
2{\frac{t}{t_0}}{\bar N}_{(t)}^k{\tilde h}_{(t)ks}dx^sdx^0+
{\frac{t^2}{t_0^2}}{\tilde h}_{(t)ks}dx^kdx^s{\Big \} }.
\ena
Here, $t_0$ is some arbitrary reference epoch (usually chosen to be the present
epoch) setting the scale of the spatial coordinates and 
${\frac{t_0}{t}}{\bar N}^k_{(t)}$ are the components of the shift vector family 
of the FOs in $({\cal N},{\bf {\bar g}}_t)$. Also, $d{\bar {\sigma}}_t^2{\equiv}
{\bar h}_{(t)ks}dx^kdx^s{\equiv}{\frac{t^2}{t_0^2}}{\bar N}_t^2{\tilde h}_{(t)ks}
dx^kdx^s$ is the spatial line element family corresponding to the spatial 
metric family ${\bf {\bar h}}_t$ intrinsic to the FHSs. Note that there are 
prior-geometric restrictions on ${\tilde h}_{(t)ks}$ (see equations (15) and 
(16) below). However, these restrictions do not show up explicitly in equation 
(1); rather, they are expressed via a certain term in one of the field 
equations (see equation (14) or (17) below). Also note that equation (1) may 
be taken as a postulate.

The time evolution of the scale factor ${\bar F}_t{\equiv}c{\bar N}_tt$ of the 
FHSs in the hypersurface-orthogonal direction may conveniently be split up
into different terms. Using the notation where a comma denotes a partial 
derivative, the symbol `${\bar {\perp}}$' denotes a scalar product with the 
negative unit normal vector field $-{\bf {\bar n}}_t$ of the FHSs, and where 
${\pounds}_{{\bf {\bar n}}_t}$ denotes a Lie derivative in the direction normal to 
the FHSs holding $t$ constant, we define
\eqa
{\bar F}_t^{-1}{\topstar{\pounds}}_{{\bf {\bar n}}_t}
{\bar F}_t{\equiv}{\bar F}_t^{-1}{\Big (}(c{\bar N}_t)^{-1}{\bar F}_t,_t+
{\pounds}_{{\bf {\bar n}}_t}{\bar F}_t{\Big )}=
{\frac{1}{c{\bar N}_tt}}+{\frac{{\bar N}_t,_t}{c{\bar N}_t^2}}
-{\frac{{\bar N}_t,_{\bar {\perp}}}{{\bar N}_t}}
{\equiv}c^{-2}{\bar x}_t+c^{-1}{\bar H}_t.
\ena
Here, $c^{-2}{\bar x}_t$ represents the {\em kinematical contribution} to 
the evolution of the spatial scale factor and $c^{-1}{\bar H}_t$ represents 
the so-called {\em non-kinematical contribution} defined by
\eqa
{\bar H}_t={\frac{1}{{\bar N}_tt}}+{\bar y}_t, \qquad
{\bar y}_t{\equiv}c^{-1}{\sqrt{{\bar a}_{{\cal F}k}{\bar a}_{\cal F}^k}}, 
\qquad c^{-2}{\bar a}_{{\cal F}j}{\equiv}{\frac{{\bar N}_t,_j}{{\bar N}_t}}.
\ena
We see from equation (3) that the non-kinematical evolution (NKE) of 
${\bar F}_t$ takes the form of an ``expansion''. Furthermore the NKE 
consists of two terms; the first term ${\frac{1}{{\bar N}_tt}}$ represents the 
{\em global NKE} of the FHSs, whereas the second term ${\bar y}_t$ represents 
the {\em local NKE} coming from the gravitational field. This second term is 
not ``realized'' globally since it is absent in equation (2). Besides, we see 
from equation (2) that the evolution of ${\bar N}_t$ with time may also be 
written as a sum of one kinematical and one non-kinematical term, i.e.,
\eqa
{\frac{{\bar N}_t,_t}{c{\bar N}_t^2}}-
{\frac{{\bar N}_t,_{\bar {\perp}}}{{\bar N}_t}}=c^{-2}{\bar x}_t+
c^{-1}{\bar y}_t. 
\ena
The split-ups defined in equations (2), (3) and (4) are necessary to be able to
construct ${\bf g}_t$ from ${\bf {\bar g}}_t$ [3]. Note that the kinematical 
evolution (KE) of the spatial scale factor may be positive or negative.

Next we define two linear, symmetric ``degenerate" connections 
${\,}{\,}{\topstar{{\bf {\bar \nabla}}}}{\,}{\,}$ and 
${\,}{\,}{\topstar{\nabla}}{\,}{\,}$ on the quasi-metric space-time manifold 
${\cal N}$. These connections are called degenerate due to the fact that they 
are essentially connections compatible with the 5-dimensional degenerate 
metrics ${\bf {\bar g}}_t$ and ${\bf g}_t$, respectively, on 
${\cal M}{\times}{\bf R}_1$ and then just restricted to $\cal N$. In the 
following, we describe the connection ${\,}{\,}{\topstar{\nabla}}{\,}{\,}$ since
this connection yields the quasi-metric equations of motion in 
$({\cal N},{\bf g}_t)$. That is, we introduce a torsion-free, metric-compatible
5-dimensional connection ${\,}{\,}{\topstar{\nabla}}{\,}{\,}$ with the property
that
\eqa
{\topstar{\nabla}}_{\frac{\partial}{{\partial}t}}{\bf g}_t=0, \qquad
{\topstar{\nabla}}_{\frac{\partial}{{\partial}t}}{\bf n}_t=0, \qquad
{\topstar{\nabla}}_{\frac{\partial}{{\partial}t}}{\bf h}_t=0,
\ena
on ${\cal M}{\times}{\bf R}_1$ and consider the restriction of 
${\,}{\,}{\topstar{\nabla}}{\,}{\,}$ to $\cal N$. Here, ${\bf h}_t$ is the
spatial metric family intrinsic to FHSs and ${\bf n}_t$ is the unit vector 
family normal to the FHSs in $({\cal N},{\bf g}_t)$. It can be shown [4] that,
expressed in a GTCS, the components which do not vanish identically of the 
degenerate connection field are given by
\eqa
{\topstar{\Gamma}}_{tj}^{{\,}i}{\equiv}
{\frac{1}{2}}h_{(t)}^{is}h_{(t)sj,t}, {\quad}
{\topstar{\Gamma}}^{\alpha}_{{\nu}{\mu}}{\equiv}{\frac{1}{2}}
g_{(t)}^{{\alpha}{\sigma}}{\Big(}g_{(t){\sigma}{\mu}},_{\nu}
+g_{(t){\nu}{\sigma}},_{\mu}-g_{(t){\nu}{\mu}},_{\sigma}{\Big)}{\equiv}
{\Gamma}^{\alpha}_{(t){\nu}{\mu}}.
\ena 
The general equations of motion for test particles are identical to the 
geodesic equation obtained from ${\,}{\,}{\topstar{\nabla}}{\,}{\,}$. In a 
GTCS they take the form (see [4] for a derivation)
\eqa
{\frac{d^2x^{\mu}}{d{\lambda}^2}}+{\Big(}
{\topstar{\Gamma}}^{\mu}_{t{\nu}}{\frac{dt}{d{\lambda}}}+
{\Gamma}^{\mu}_{(t){\beta}{\nu}}{\frac{dx^{\beta}}{d{\lambda}}}{\Big)}
{\frac{dx^{\nu}}{d{\lambda}}} 
={\Big(}{\frac{cd{\tau}_t}{d{\lambda}}}{\Big)}^2c^{-2}a_{(t)}^{\mu},
\ena
where $d{\tau}_t$ is the proper time as measured along the curve, ${\lambda}$ 
is some general affine parameter and ${\bf a}_t$ is the 4-acceleration as 
measured along the curve. From equations (6) and (7), we see that quasi-metric 
theory cannot be identified with any metric theory since the affine connection 
compatible with a general metric family is non-metric.

As mentioned above, a basic property of the QMF is that gravitational 
quantities will be formally variable when measured in atomic units. In 
particular, this applies to the ``bare'' gravitational coupling parameter 
$G^{\rm B}_t$, formally varying like length squared when measured in atomic 
units (i.e., like ${\bar F}_t^2$). Now $G^{\rm B}_t$ couples to charge squared, 
or more generally to the electromagnetic stress-energy tensor [5]. On the other 
hand, for material sources, masses formally vary as ${\bar F}_t^{-1}$, but this 
is not measurable in non-gravitational experiments. This means that the 
``screened'' gravitational parameter $G^{\rm S}_t$ measured for material sources
effectively varies as ${\bar F}_t$. Consequently, local gravitational 
experiments designed to measure gravitational coupling parameters should depend
on source composition, so that it will be necessary to distinguish between 
$G^{\rm B}_t$ and $G^{\rm S}_t$.

However, it is convenient to {\em define} constants $G^{\rm B}$ and $G^{\rm S}$ 
as the values of $G^{\rm B}_t$ and $G^{\rm S}_t$, respectively, measured in 
(hypothetical) local gravitational experiments at the arbitrary reference epoch
$t_0$, such that the formal variabilities of $G^{\rm B}_t$ and $G^{\rm S}_t$ are 
transferred to mass (and charge, if any). Thus, we have to distinguish
between {\em active mass}, which is a scalar field, and {\em passive mass} 
(passive gravitational mass and inertial mass). (Similarly one must distinguish
between {\em active charge} and {\em passive charge} [5].) For a material 
particle, the above discussion implies that active mass $m_t$ varies formally 
as ${\bar F}_t$ measured in atomic units (but passive mass does of course not 
vary). That is, for a material particle we have that
\eqa
m_t,_t={\Big (}{\frac{1}{t}}+{\frac{{\bar N}_t,_t}{{\bar N}_t}}{\Big )}m_t, 
{\qquad}
m_t,_{\bar {\perp}}={\frac{{\bar N}_t,_{\bar {\perp}}}{{\bar N}_t}}m_t, \qquad
m_t,_j=c^{-2}{\bar a}_{{\cal F}j}m_t,
\ena 
where ${\bf {\bar a}}_{\cal F}$ is the 4-acceleration of the FOs in the family 
${\bf {\bar g}}_t$. On the other hand, for a local electromagnetic source,
active mass $m_t$ (or active energy) varies formally as ${\bar F}_t^2$ measured
in atomic units. (For extended electromagnetic sources, one must also take into 
account a secular attuenation (not noticeable locally) of the electromagnetic 
field [3, 5]). For the rest of the present paper we will assume no net charge 
but that photons as a gravitational source cannot always be neglected.

Taking into the account said variation of active mass in quasi-metric 
space-time, it is possible to find local conservation laws. These local 
conservation laws are valid for fixed $t$ and involve the metric covariant 
divergence ${\bf {\bar \nabla}}_t{\cdot}{\bf T}_t$ (using the Levi-Civita 
connection ${\bf {\bar \nabla}}_t$) of the active stress-energy tensor 
${\bf T}_t$. They take the form (in component notation) [4]
\eqa
T_{(t){\mu};{\nu}}^{\nu}=2{\frac{{\bar N}_t,_{\nu}}{{\bar N}_t}}
T_{(t){\mu}}^{\nu}.
\ena
If the dependence on $t$ of ${\bf T}_t$ is entirely due to the above mentioned
formal variability, ${\bf T}_t$ is locally conserved when $t$ varies as well.
{\em These local conservation laws are valid independent of the nature of
the gravitating source}, i.e., they are valid for material sources as well as 
for electromagnetic sources. Note that local conservation of ${\bf T}_t$ 
implies that inertial observers move along geodesics of ${\,}{\,}
{\topstar{\bf {\bar{ \nabla}}}}{\,}{\,}$ in $({\cal N},{\bf {\bar g}}_t)$, and 
that this guarantees that inertial observers move along geodesics of of 
${\,}{\,}{\topstar{\bf {\nabla}}}{\,}{\,}$ in $({\cal N},{\bf g}_t)$ as well 
[3, 4]. This means that the equations of motion (7) are consistent with the 
local conservation laws (9).

It is useful to project these local conservation laws with respect to the 
FHSs. We then get the equations (in content equivalent to equation (9))
\eqa
{\cal L}_{{\bf {\bar n}}_t}T_{(t){\bar {\perp}}{\bar {\perp}}}=
{\Big (}{\bar K}_t-2{\frac{{\bar N}_t,_{\bar {\perp}}}{{\bar N}_t}}
{\Big )}T_{(t){\bar {\perp}}{\bar {\perp}}}
+{\bar K}_{(t)ik}{\hat T}_{(t)}^{ik}-{\hat T}^i_{(t){\bar {\perp}}{\mid}i},
\ena
\eqa
{\frac{1}{{\bar N}_t}}{\cal L}_{{\bar N}_t{\bf {\bar n}}_t}
T_{(t)j{\bar {\perp}}}={\Big (}{\bar K}_t
-2{\frac{{\bar N}_t,_{\bar {\perp}}}{{\bar N}_t}}
{\Big )}T_{(t)j{\bar {\perp}}}-c^{-2}{\bar a}_{{\cal F}j}
T_{(t){\bar {\perp}}{\bar {\perp}}}
+c^{-2}{\bar a}_{{\cal F}i}{\hat T}_{(t)j}^i-{\hat T}^i_{(t)j{\mid}i},
\ena
where ${\cal L}_{{\bf {\bar n}}_t}$ denotes a Lie derivative of spatial objects in 
the direction normal to the FHSs (with $t$ fixed). (A ``hat'' denotes an object
intrinsic to the FHSs.) See [4] for a derivation of these equations. Also 
postulated in [4] are the quasi-metric field equations involving the geometric 
tensor family ${\bf {\bar Q}}_t$, the active electromagnetic stress-energy 
tensor ${\bf T}^{\rm (EM)}_t$ and the active stress-energy-tensor for material 
sources ${\bf T}^{\rm (MA)}_t$ (where $`{\mid}$' denotes a
space covariant derivative and where
${\kappa}^{\rm B}{\equiv}{\frac{8{\pi}G^{\rm B}}{c^4}}$,
${\kappa}^{\rm S}{\equiv}{\frac{8{\pi}G^{\rm S}}{c^4}}$)
\eqa
{\bar Q}_{(t){\bar {\perp}}{\bar {\perp}}}{\equiv}
2{\bar R}_{(t){\bar {\perp}}{\bar {\perp}}}=
2(c^{-4}{\bar a}_{{\cal F}k}{\bar a}_{\cal F}^k+
c^{-2}{\bar a}^k_{{\cal F}{\mid}k}-{\bar K}_{(t)ik}{\bar K}_{(t)}^{ik}
+{\cal L}_{{\bf {\bar n}}_t}{\bar K}_t) \nonumber \\
={\kappa}^{\rm B}(T^{\rm (EM)}_{(t){\bar {\perp}}{\bar {\perp}}}+
{\hat T}_{(t)i}^{{\rm (EM)}i})+
{\kappa}^{\rm S}(T^{\rm (MA)}_{(t){\bar {\perp}}{\bar {\perp}}}+
{\hat T}_{(t)i}^{{\rm (MA)}i}),
\ena
\eqa
{\bar Q}_{(t)j{\bar {\perp}}}{\equiv}{\bar K}_{(t)j{\mid}i}^i-{\bar K}_t,_j
+{\Big (}{\frac{{\bar h}_{(t)}^{ik}}{{\bar N}_t}}
{\frac{\partial}{{\partial}x^0}}{\bar h}_{(t)ij}{\Big )}_{{\mid}k}-
{\Big (}{\frac{{\bar h}_{(t)}^{ik}}{{\bar N}_t}}
{\frac{\partial}{{\partial}x^0}}{\bar h}_{(t)ik}{\Big )},_j
={\kappa}^{\rm B}T^{{\rm (EM)}}_{(t)j{\bar {\perp}}}
+{\kappa}^{\rm S}T^{\rm (MA)}_{(t)j{\bar {\perp}}}.
\ena
Here, ${\bf {\bar R}}_t$ is the Ricci tensor family, ${\bf {\bar G}}_t$ is the
Einstein tensor family and ${\bf {\bar K}}_t$ is the extrinsic curvature tensor
family of the FHSs corresponding to the metric family (1). (${\bar K}_t$ is the
trace of ${\bf {\bar K}}_t$.) The set of quasi-metric field equations is 
completed with the traceless quantity
\eqa
{\bar Q}_{(t)ij}{\equiv}{\frac{1}{{\bar N}_t}}{\cal L}_{{\bar N}_t
{\bf {\bar n}}_t}{\bar K}_{(t)ij}+
{\frac{1}{3}}{\Big [}2{\bar K}_{(t)ks}{\bar K}_{(t)}^{ks}-{\bar K}_t^2
-{\cal L}_{{\bf {\bar n}}_t}{\bar K}_t
{\Big ]}{\bar h}_{(t)ij}+{\bar K}_t{\bar K}_{(t)ij}
\nonumber \\
-c^{-2}{\bar a}_{{\cal F}i{\mid}j}-
c^{-4}{\bar a}_{{\cal F}i}{\bar a}_{{\cal F}j}
+{\Big [}c^{-2}{\bar a}_{{\cal F}{\mid}s}^s
-{\frac{1}{(ct{\bar N}_t)^2}}{\Big ]}{\bar h}_{(t)ij}-{\bar H}_{(t)ij}=0,
\ena
where the prior-geometric requirement on the spatial Ricci curvature scalar 
family ${\bar P}_t$,
\eqa
{\bar P}_t=-4c^{-2}{\bar a}_{{\cal F}{\mid}s}^s
+2c^{-4}{\bar a}_{{\cal F}}^s{\bar a}_{{\cal F}s}+{\frac{6}{(ct{\bar N}_t)^2}},
\ena
ensures that equation (14) is indeed manifestly traceless. Besides, the 
components of the spatial Einstein tensor family ${\bf {\bar H}}_t$ is given by
\eqa
{\bar H}_{(t)ij}=-c^{-2}{\bar a}_{{\cal F}i{\mid}j}-
c^{-4}{\bar a}_{{\cal F}i}{\bar a}_{{\cal F}j}+
c^{-2}{\bar a}_{{\cal F}{\mid}s}^s{\bar h}_{(t)ij}+{\tilde H}_{(t)ij},
\ena
where ${\bf {\tilde H}}_t$ is the spatial Einstein tensor family calculated 
from the spatial metric family ${\bf {\tilde h}}_t{\equiv}
{\frac{t_0^2}{t^2}}{\bar N}_t^{-2}{\bf {\bar h}}_t$ in 
$({\cal N},{\bf {\bar g}}_t)$. Note that, while equation (16) implies that 
${\tilde P}_t={\frac{6}{(ct_0)^2}}$ is fixed by the prior 3-geometry, we have
that ${\tilde H}_{(t)ij}$ is not necessarily equal to the prior-geometric 
quantity $-{\frac{1}{(ct_0)^2}}{\tilde h}_{(t)ij}$. This shows that, while there 
is prior 3-geometry, there is still sufficient dynamical freedom left 
associated with the metric family ${\bf {\tilde h}}_t$. This is 
further illustrated by writing equation (14) in the form (using equations
(12) and (16))
\eqa
{\frac{1}{{\bar N}_t}}{\cal L}_{{\bar N}_t{\bf {\bar n}}_t}{\bar K}_{(t)ij}
+{\bar K}_t{\bar K}_{(t)ij}-{\tilde H}_{(t)ij} \nonumber \\
={\frac{1}{3}}{\Big [}{\bar R}_{(t){\bar {\perp}}{\bar {\perp}}}
+{\bar K}_t^2-{\bar K}_{(t)ks}{\bar K}_{(t)}^{ks}
-c^{-2}{\bar a}_{{\cal F}{\mid}s}^s
-c^{-4}{\bar a}_{{\cal F}}^s{\bar a}_{{\cal F}s}+{\frac{3}{(ct{\bar N}_t)^2}}
{\Big ]}{\bar h}_{(t)ij}.
\ena
We notice that by taking the trace of equation (17), we recover the (general) 
expression for ${\bar R}_{(t){\bar {\perp}}{\bar {\perp}}}$ given in equation (12).
Besides, note that equation (17) is only partially coupled to matter sources
via the the scalar quantity ${\bar R}_{(t){\bar {\perp}}{\bar {\perp}}}$ and equation 
(12). That is, equation (17) (or equation (14)) is not fully coupled to the 
corresponding projection of ${\bf T}_t$, so the above field equations only 
represent a {\em partial} coupling of ${\bf {\bar Q}}_t$ to ${\bf T}_t$. 
Nevertheless, due to the form of equation (17) (or equation (14)), just as for 
General Relativity, the quasi-metric field equations yield two independent 
propagating dynamical degrees of freedom. Also note that ${\bf {\bar Q}}_t$ 
is not a ``genuine'' space-time tensor family since it is {\em defined} from 
its projections with respect to a particular foliation of quasi-metric 
space-time into spatial hypersurfaces (i.e., the FHSs). See [3, 4] for a 
further discussion. Finally we notice that equation (14) follows from the 
criterion [3, 4]
\eqa
{\bar C}_{(t){\bar {\perp}}i{\bar {\perp}}j}={\tilde H}_{(t)ij}+
{\frac{1}{(ct{\bar N}_t)^2}}{\bar h}_{(t)ij},
\ena
involving a particular projection of the Weyl tensor family ${\bf {\bar C}}_t$.

It should be emphasized that, although the field equations are postulated 
rather than derived, they are by no means arbitrary; equation (12) for example,
follows naturally from a geometrical correspondence with Newton-Cartan theory 
and has a similar (apart from the composition-dependent coupling) counterpart 
valid for General Relativity. Besides, equation (15) implies that the temporal 
evolution of the FHSs does not have any gauge freedom associated with it, so 
that as opposed to canonical General Relativity, there is no freedom to choose 
lapse and shift. This means that the quasi-metric field equations determine the
FHSs uniquely as well as the metric family ${\bf {\bar g}}_t$, and that 
equations (12)-(17) are exactly valid only for projections with respect to the 
FHSs.

The uniqueness of $t$ and thus the foliation of ${\bf {\bar g}}_t$ into a 
unique set of FHSs is a result of the topology of quasi-metric space-time where
space is by definition compact and with positive global curvature scalar
${\tilde P}_t$. Said topology makes it natural to identify the FOs with 
observers being approximately at rest with respect to the cosmic rest frame 
associated with the smeared-out motion of the galaxies (i.e., the frame where 
the cosmic relic microwave background is observed to be isotropic on average). 
This identification is practical when doing quasi-metric cosmology but not when 
doing gravitational physics for isolated systems. However, for a sufficiently 
small isolated system we may set ${\tilde P}_t{\approx}0$ so that the FHSs may 
be treated as approximately asymptotically flat. Then an alternative 
(approximately global) time function $t'$ foliating ${\bf {\bar g}}_{t'}$ into 
an alternative set of hypersurfaces (also being asymptotically flat) may be 
defined. That is, an alternative class of observers always moving orthogonally 
to the alternative hypersurfaces may be defined such that said observers are 
at rest with respect to the barycentre of the isolated system. Moreover, the 
field equations (with ${\tilde P}_{t'}=0$) may then be transformed with respect
to this new set of hypersurfaces. However, the field equations will not be 
invariant under said transformation; rather they will depend on the velocity of
the isolated system with respect to the cosmic rest frame. But in practice, the 
``preferred frame''-effects introduced by said procedure should be small (at 
most of post-Newtonian order), if the size of the isolated system is small 
compared to $ct_0$ and its local speed with respect to the cosmic rest frame is
much smaller than the speed of light.

In what follows, we shall do calculations applied to some so-called 
``metrically static'' gravitational systems (see the next section for an 
explanation). Then it is convenient to have a specific
expression for the geometry intrinsic to 
the FHSs obtained from equation (16). That is, for these cases
${\bf {\bar K}}_t$ vanishes identically, so using equation (14) we find that
\eqa
{\bar H}_{(t)ij}=c^{-2}{\Big (}{\bar a}^k_{{\cal F}{\mid}k}-
{\frac{1}{({\bar N}_tt)^2}}{\Big )}{\bar h}_{(t)ij}
-c^{-4}{\bar a}_{{\cal F}i}{\bar a}_{{\cal F}j}
-c^{-2}{\bar a}_{{\cal F}i{\mid}j}.
\ena
We notice that also this expression includes a prior-geometric term.
\subsection{Special equations of motion}
In this paper, we analyse the equations of motion (7) in the case of a 
uniformly expanding, isotropic gravitational field in vacuum exterior to an 
isolated, spherically symmetric source in an isotropic, compact spatial 
background. We also require that the source is at rest with respect to some 
GTCS. Furthermore, we require that ${\bar N}_t$ and ${\tilde h}_{(t)ks}$ are
independent of $x^0$ and $t$; i.e., that the only explicit time dependence 
is via $t$ in the spatial scale factor (using the chosen GTCS). Then it turns 
out that also the FOs must be at rest with respect to the chosen GTCS and 
consequently the shift vector field vanishes (this is strictly true only if
the isolated system is at rest with respect to the cosmic rest frame, but see
the previous section for a discussion of why this is not a crucial issue). We 
denote this a ``metrically static'' case. This scenario may be taken as a 
generalization of the analogous case with a Minkowski background (that case is 
analysed in [4]) and is more realistic since the Minkowski background is not a 
part of our theory but rather invoked as an approximation being useful in 
particular cases.

We start by making a specific {\em ansatz} for the form of ${\bf {\bar g}}_t$.
Introducing a spherical GTCS ${\{}x^0,r,{\theta},{\phi}{\}}$, 
where $r$ is a Schwarzschild radial coordinate, we assume that the 
metric families ${\bf {\bar g}}_t$ and ${\bf g}_t$ can be written in a form 
compatible with equations (1) and (19) (using the notation 
$'{\hspace*{2mm}}{\equiv}{\frac{\partial}{{\partial}r}}$), i.e.,
\eqa
c^2{\overline{d{\tau}}}^2_t={\bar{B}}(r)(dx^0)^2-({\frac{t}{t_0}})^2{\Big(}
{\bar{A}}(r)dr^2+r^2d{\Omega}^2{\Big)}, {\nonumber} \\
c^2d{\tau}^2_t=B(r)(dx^0)^2-({\frac{t}{t_0}})^2{\Big(}
A(r)dr^2+r^2d{\Omega}^2{\Big)},
\ena
\eqa
{\bar A}(r){\equiv}{\frac{{\Big [}1-{\frac{r}{2}}
{\frac{{\bar B}'(r)}{{\bar B}(r)}}
{\Big ]}^2}{1-{\frac{r^2}{{\bar B}(r){\Xi}^2_0}}}},
\ena
where ${\bar B}(r){\equiv}{\bar N}_t^2(r)$,
$d{\Omega}^2{\equiv}d{\theta}^2+{\sin}^2{\theta}d{\phi}^2$ and
${\Xi}_0{\equiv}ct_0$. We notice that for metrically static cases, we have from
equation (19) that ${\tilde h}_{(t)ks}=S_{ks}$, where $S_{ks}dx^kdx^s$ is the 
metric of the 3-sphere ${\bf S}^3$ (with radius equal to $ct_0$). This is why
the function ${\bar{A}}(r)$  takes the form (21). Moreover, we notice that the 
spatial coordinate system covers only half of ${\bf S}^3$, thus the range of 
the radial coordinate is $r{\leq}{\Xi}_0$ only. The function ${\bar{B}}(r)$ 
and thus the function ${\bar{A}}(r)$ may be calculated from the field 
equations; we treat this problem in the next section. The functions $A(r)$ and 
$B(r)$ may then be found from ${\bf {\bar g}}_t$ and ${\bar y}_t$.

We now calculate the metric connection coefficients from the metric family 
${\bf g}_t$ given in equation (20). A straightforward calculation yields 
\eqa
{\Gamma}^r_{(t)rr}&=&{\frac{A'(r)}{2A(r)}}, {\qquad}
{\Gamma}^r_{(t){\theta}{\theta}}=-{\frac{r}{A(r)}}, {\qquad}
{\Gamma}^r_{(t){\phi}{\phi}}={\Gamma}^r_{(t){\theta}{\theta}}{\sin}^2{\theta},
{\nonumber} \\
{\Gamma}^r_{(t)00}&=&({\frac{t_0}{t}})^2{\frac{B'(r)}{2A(r)}}, 
{\qquad}{\Gamma}^{\theta}_{(t)r{\theta}}={\Gamma}^{\theta}_{(t){\theta}r}=
{\frac{1}{r}}, {\qquad} {\Gamma}^{\theta}_{(t){\phi}{\phi}}=
-{\sin}{\theta}{\cos}{\theta}, {\nonumber} \\
{\Gamma}^{\phi}_{(t)r{\phi}}&=&{\Gamma}^{\phi}_{(t){\phi}r}={\frac{1}{r}},
{\qquad}{\Gamma}^{\phi}_{(t){\phi}{\theta}}=
{\Gamma}^{\phi}_{(t){\theta}{\phi}}={\cot}{\theta}, {\qquad}
{\Gamma}^0_{(t)0r}={\Gamma}^0_{(t)r0}={\frac{B'(r)}{2B(r)}}.
\ena
In the following we use the equations of motion (7) to find the paths 
of inertial test particles moving in the metric family ${\bf g}_t$. Since 
${\bf a}_t$ vanishes for inertial test particles, we get the relevant equations 
by using equation (6) and inserting the expressions (22) into equation (7). 
This yields (making explicit use of the fact that $cdt=dx^0$ in a GTCS)
\eqa
{\frac{d^{2}r}{d{\lambda}^{2}}} + 
{\frac{A'(r)}{2A(r)}}{\Big(}{\frac{dr}{d{\lambda}}}{\Big)}^2 -
{\frac{r}{A(r)}}{\bigg[}{\Big(}{\frac{d{\theta}}{d{\lambda}}}{\Big)}^{2} + 
{\sin}^{2}{\theta}{\Big(}{\frac{d{\phi}}{d{\lambda}}}{\Big)}^{2}{\bigg]}
{\nonumber} \\
+({\frac{t_0}{t}})^2
{\frac{B'(r)}{2A(r)}}{\Big(}{\frac{dx^0}{d{\lambda}}}{\Big)}^{2} + 
{\frac{1}{ct}}{\frac{dr}{d{\lambda}}}{\frac{dx^0}{d{\lambda}}}=0,
\ena
\eqa
{\frac{d^{2}{\theta}}{d{\lambda}^{2}}} +
{\frac{2}{r}}{\frac{d{\theta}}{d{\lambda}}}{\frac{dr}{d{\lambda}}} -
{\sin}{\theta}{\cos}{\theta}{\Big(}{\frac{d{\phi}}{d{\lambda}}}{\Big)}^{2} + 
{\frac{1}{ct}}{\frac{d{\theta}}{d{\lambda}}}{\frac{dx^0}{d{\lambda}}} = 0,
\ena
\eqa
{\frac{d^{2}{\phi}}{d{\lambda}^{2}}} +
{\frac{2}{r}}{\frac{d{\phi}}{d{\lambda}}}{\frac{dr}{d{\lambda}}} +
2{\cot}{\theta}{\frac{d{\phi}}{d{\lambda}}}{\frac{d{\theta}}{d{\lambda}}} + 
{\frac{1}{ct}}{\frac{d{\phi}}{d{\lambda}}}{\frac{dx^0}{d{\lambda}}} = 0,
\ena
\eqa
{\frac{d^{2}x^0}{d{\lambda}^{2}}} +
{\frac{B'(r)}{B(r)}}{\frac{dx^0}{d{\lambda}}}{\frac{dr}{d{\lambda}}} =0.
\ena
If we restrict the motion to the equatorial plane, equation (24) becomes 
vacant, and equation (25) reduces to
\eqa
{\frac{d^{2}{\phi}}{d{\lambda}^{2}}} +
{\frac{2}{r}}{\frac{d{\phi}}{d{\lambda}}}{\frac{dr}{d{\lambda}}} +
{\frac{1}{ct}}{\frac{d{\phi}}{d{\lambda}}}{\frac{dx^0}{d{\lambda}}}=0.
\ena
Dividing equation (27) by ${\frac{d{\phi}}{d{\lambda}}}$ we find (assuming
${\frac{d{\phi}}{d{\lambda}}}{\neq}0$)
\eqa
{\frac{d}{d{\lambda}}}{\bigg[}{\ln}{\Big(}{\frac{d{\phi}}{d{\lambda}}}{\Big)} 
+ {\ln}{\Big(}r^{2}{\frac{t}{t_0}}{\Big)}{\bigg]}=0.
\ena
We thus have a constant of the motion, namely
\eqa
J{\equiv}{\frac{t}{t_0}}r^2{\frac{d{\phi}}{d{\lambda}}}.
\ena
Dividing equation (26) by ${\frac{dx^0}{d{\lambda}}}$ yields
\eqa
{\frac{d}{d{\lambda}}}{\bigg[}{\ln}{\Big(}{\frac{dx^0}{d{\lambda}}}{\Big)} + 
{\ln}B(r){\bigg]}=0.
\ena
Equation (30) yields a constant of the motion which we can absorb
into the definition of ${\lambda}$ such that a solution of equation
(30) is [7]
\eqa
{\frac{dx^0}{d{\lambda}}}={\frac{1}{B(r)}}.
\ena
Multiplying equation (23) by 
${\frac{2t^2A(r)}{t_0^2}}{\frac{dr}{d{\lambda}}}$ and 
using the expressions (29), (31) we find that
\eqa
{\frac{d}{d{\lambda}}}{\bigg[}
{\frac{t^2A(r)}{t_0^2}}{\Big(}{\frac{dr}{d{\lambda}}}{\Big)}^2-
{\frac{1}{B(r)}}+{\frac{J^2}{r^2}}{\bigg]}=0,
\ena
thus a constant $E$ of the motion is defined by
\eqa
{\frac{t^2A(r)}{t_0^2}}{\Big(}{\frac{dr}{d{\lambda}}}{\Big)}^2-
{\frac{1}{B(r)}}+{\frac{J^2}{r^2}}{\equiv}-E.
\ena
Equation (33) may be compared to an analogous expression obtained for the 
spherically symmetric, static gravitational field in the metric framework [7].
Inserting the formulae (29), (31) and (33) into equation (20) and using the 
fact that in a GTCS we can formally write $dx^0=cdt$ when traversing the 
family of metrics, we find that
\eqa
c^2d{\tau}_t^2=Ed{\lambda}^2.
\ena
Thus our equations of motion (7) force $d{\tau}_t/d{\lambda}$ to be
constant, quite similarly to the case when the total connection is
metric, as in the metric framework. From equation (34) we see that we must 
have $E=0$ for photons and $E>0$ for material particles.

We may eliminate the parameter ${\lambda}$ from equations (29), (31), (33) 
and (34) and alternatively use $t$ as a time parameter. This yields
\eqa
{\frac{t}{t_0}}r^2{\frac{d{\phi}}{cdt}}=B(r)J,
\ena
\eqa
({\frac{t}{t_0}})^2A(r)B^{-2}(r){\Big(}{\frac{dr}{cdt}}{\Big)}^2-
{\frac{1}{B(r)}}+{\frac{J^2}{r^2}}{\equiv}-E,
\ena
\eqa
d{\tau}_t^2=EB^2(r)dt^2.
\ena
We may integrate equations (35) and (36) to find the time history 
$(r(t),{\phi}(t))$ along the curve if the functions $A(r)$ and $B(r)$
are known.

For the spherically symmetric, static vacuum metric case with no global NKE, 
one can solve the geodesic equation for particles orbiting in circles with 
different radii, and from this find the asymptotically Keplerian nature of the 
corresponding rotational curve [7]. In our case we see from equations (35) and 
(36) that we can find circle orbits as solutions; such orbits have the 
property that the orbital speed 
$w(r)=B^{-1/2}(r){\frac{t}{t_0}}r{\frac{d{\phi}}{dt}}$ is independent of $t$. 
(Here $w(r)$ is the norm of the 3-velocity
${\bf w}_t={\sqrt{1-{\frac{w^2}{c^2}}}}{\frac{d{\phi}}{d{\tau}_t}}
{\frac{\partial}{{\partial}{\phi}}}$.)
\section{Metrically static, spherically symmetric systems}
\subsection{Perfect fluid sources}
We now seek general solutions of the type (20) of the field equations, and
where the source is modelled as a perfect fluid. Then any active stress-energy
tensor ${\bf T}_t$ takes the form (valid for both a photon fluid and 
material perfect fluid sources)
\eqa
{\bf T}_t=({\tilde{\varrho}}_{\text m}+{\tilde p}/c^2){\bf {\bar u}}_t{\otimes}
{\bf {\bar u}}_t+{\tilde p}{\bf {\bar g}}_t,
\ena
where ${\tilde {\varrho}}_{\text m}$ is the density of active mass-energy and 
${\tilde p}$ is the active pressure seen in the local rest frame of the 
fluid. Moreover, ${\bf {\bar u}}_t$ is the 4-velocity of the fluid in 
$({\cal N},{\bf {\bar g}}_t)$. But what can be measured locally is not 
${\bf T}_t$ but {\em the passive stress-energy tensor} 
${\bf {\bar {\cal T}}}_t$ in $({\cal N},{\bf {\bar g}}_t)$, given by
\eqa
{\bf {\bar {\cal T}}}_t=({\varrho}_{\text m}+p/c^2)
{\bf {\bar u}}_t{\otimes}{\bf {\bar u}}_t+p{\bf {\bar g}}_t,
\ena
where ${\varrho}_{\text m}$ is the passive mass-energy as measured in the local
rest frame of the fluid and $p$ is the associated passive pressure. Observe 
that the counterpart ${\cal T}_t$ in $({\cal N},{\bf g}_t)$ to
${\bf {\bar {\cal T}}}_t$ is given by
\eqa
{\cal T}_t={\sqrt{\frac{{\bar h}_t}{h_t}}}{\Big [}({\varrho}_{\text m}+p/c^2)
{\bf u}_t{\otimes}{\bf u}_t+p{\bf g}_t{\Big ]},
\ena
where ${\bar h}_t$ and $h_t$ are the determinants of the spatial metrics
${\bf {\bar h}}_t$ and ${\bf h}_t$, respectively. Since electromagnetic and 
material active mass-energy have different formal variability in quasi-metric 
space-time, we have that the relationship between ${\tilde {\varrho}}_{\text m}$ 
and ${\varrho}_{\text m}$ is given by
\eqa
{\tilde {\varrho}}_{\text m}=
\left\{
\begin{array}{ll}
{\frac{t}{t_0}}{\bar N}_t{\varrho}_{\text m}
& \text{for a fluid of material particles,} \\ [1.5ex]
{\frac{t^2}{t_0^2}}{\bar N}_t^2{\varrho}_{\text m} & \text{for the 
electromagnetic field,}
\end{array}
\right.
\ena
and similarly for the relationship between ${\tilde p}$ and $p$. In the 
following sections, we set up the relevant equations both for the interior and 
the exterior gravitational field. As we shall see, it is possible to find
an implicit exact solution inside the source but this solution is not very 
useful. Therefore, to find explicit solutions for the interior field, the 
equations should be solved numerically. However, we have not performed any 
numerical calculations. On the other hand an explicit exact solution may be 
found for the exterior field.
\subsection{The interior field}
In this section we analyse the gravitational field inside the source. That is,
we do the necessary analytical calculations in order to write the relevant
equations in a form appropriate for numerical treatment. For this purpose, it 
is convenient to rewrite the line element ${\overline{ds}}^2_t$ given in 
equation (20) by changing to a new radial coordinate
${\rho}{\equiv}r/{\sqrt{{\bar B}}}$. Then we find, using equation (21), that
\eqa
{\overline{ds}}^2_t={\bar{B}}({\rho}){\Big [}-(dx^0)^2+({\frac{t}{t_0}})^2
{\Big(}
{\frac{d{\rho}^2}{1-{\frac{{\rho}^2}{{\Xi_0^2}}}}}
+{\rho}^2d{\Omega}^2{\Big)}{\Big ]}.
\ena
Furthermore, from equation (42) and the definitions, we find that
\eqa
c^{-2}{\bar a}_{{\cal F}{\rho}}={\frac{{\bar B},_{\rho}}{2{\bar B}}}, {\quad}
c^{-2}{\bar a}_{{\cal F}{\rho}{\mid}{\rho}}=
{\frac{{\bar B},_{{\rho}{\rho}}}{2{\bar B}}}-
{\frac{3}{4}}{\Big (}{\frac{{\bar B},_{\rho}}{{\bar B}}}{\Big )}^2-
{\frac{{\rho}}{{\Xi}_0^2(1-{\frac{{\rho}^2}{{\Xi_0^2}}})}}
{\frac{{\bar B},_{\rho}}{2{\bar B}}}, 
\nonumber \\
c^{-2}{\sin}^{-2}{\theta}{\bar a}_{{\cal F}{\phi}{\mid}{\phi}}=
c^{-2}{\bar a}_{{\cal F}{\theta}{\mid}{\theta}}={\frac{{\rho}^2}{2}}
(1-{\frac{{\rho}^2}{{\Xi_0^2}}}){\Big [}{\frac{1}{2}}
{\Big (}{\frac{{\bar B},_{\rho}}{{\bar B}}}{\Big )}^2
+{\frac{1}{\rho}}{\frac{{\bar B},_{\rho}}{{\bar B}}}{\Big ]},
{\nonumber} \\
c^{-2}{\bar a}^k_{{\cal F}{\mid}k}=({\frac{t_0}{t}})^2{\frac{1}{\bar B}}
{\Big {\{ }}(1-{\frac{{\rho}^2}{{\Xi_0^2}}}){\Big [}
{\frac{{\bar B},_{{\rho}{\rho}}}{2{\bar B}}}-
{\frac{1}{4}}{\Big (}{\frac{{\bar B},_{\rho}}{{\bar B}}}{\Big )}^2{\Big ]}
+{\frac{1}{\rho}}(1-{\frac{3{\rho}^2}{2{\Xi_0^2}}})
{\frac{{\bar B},_{\rho}}{{\bar B}}}{\Big {\}}}.
\ena
For a fluid of material particles, active mass density varies formally as 
${\bar F}_t^{-2}$ whereas for electromagnetic field energy (e.g., photon 
energy), it varies as ${\bar F}_t^{-1}$ according to equation (41). However, 
the metrically static condition requires that ${\bar B}$ must be independent 
of $t$, implying that the time variability of source densities must cancel out 
in the field equations. This means that we must require a cosmic redshift of 
photon energy, yielding an extra factor ${\frac{t_0}{t}}$ in the source photon 
energy density. Besides, gravitational spectral shifts of photon energy must 
yield an extra factor ${\bar N}_t^{-1}$, so that one effectively gets an extra 
factor ${\bar F}_t^{-1}$ for photon sources. Thus, for a metrically static 
source, we may treat material particle sources and photons equally, as if 
active mass of both formally vary as ${\bar F}_t^{-2}$ (this approximation is 
only valid if the net energy transfer between photons and material particles 
is negligible).

For reasons of convenience, we choose to extract this formal variability 
explicitly. What is left after separating out the formal variability from the 
active mass density, is by definition {\em the properly scaled density 
of active mass} ${\bar {\varrho}}_{\text m}{\equiv}
{\bar {\varrho}}^{\rm (EM)}_{\text m}+{\bar {\varrho}}^{\rm (MA)}_{\text m}$. (The 
corresponding pressure is ${\bar p}$.) For the metrically static case 
$T_{(t){\bar {\perp}}j}=0$, and besides we find from equation (38) that
\eqa
T_{(t){\bar {\perp}}{\bar {\perp}}}={\tilde {\varrho}}_{\text m}c^2{\equiv}
{\frac{t_0^2}{t^2}}{\frac{{\bar {\varrho}}_{\text m}c^2}{{\bar B}}}{\equiv}
{\frac{t_0^2}{t^2{\bar B}}}[{\bar {\varrho}}^{\rm (EM)}_{\text m}c^2+
{\bar {\varrho}}^{\rm (MA)}_{\text m}c^2], \nonumber \\ 
T_{(t)r}^r=T_{(t){\theta}}^{\theta}=T_{(t){\phi}}^{\phi}={\tilde p}{\equiv}
{\frac{t_0^2}{t^2}}{\frac{{\bar p}}{{\bar B}}}{\equiv}
{\frac{t_0^2}{t^2{\bar B}}}[{\bar p}^{\rm (EM)}_{\text m}+
{\bar p}^{\rm (MA)}_{\text m}],
\ena
where ${\bar {\varrho}}_{\text m}$ and ${\bar p}$ do not depend on $t$
since the direct effects of the cosmic expansion have been scaled out. 
We thus have, by using equations (44) and the local conservation laws (10), 
(11) applied to the metrically static case, that
(with ${\dot {\ }}{\equiv}{\frac{\partial}{{\partial}t}}$)
\eqa
{\dot {\bar {\varrho}}}_{\text m}={\dot {\bar p}}=0, \qquad
{\bar p},_{\rho}^{\rm (EM)}={\frac{{\bar p}^{\rm (EM)}}{\bar p}}
{\bar p},_{\rho}, \qquad {\bar p},_{\rho}^{\rm (MA)}=
{\frac{{\bar p}^{\rm (MA)}}{\bar p}}{\bar p},_{\rho}, \nonumber \\
{\bar p},_{\rho}=-c^{-2}{\bar a}_{{\cal F}{\rho}}({\bar {\varrho}}_{\text m}c^2
-3{\bar p})=-({\bar {\varrho}}^{\rm (MA)}_{\text m}c^2-3{\bar p}^{\rm (MA)})
{\frac{{\bar B},_{\rho}}{2{\bar B}}},
\ena
where the last equation holds since the only contribution to 
${\bf T}^{\rm (EM)}_t$ comes from electromagnetic radiation. This means that 
radiation does not contribute to ${\bar p},_{\rho}$.

Equations (45) are valid for any metrically static perfect fluid. But to have 
experimental input, we need to specify an equation of state 
$p=p({\varrho}_{\text m})$ consistent with the metrically static condition. That 
is, the metrically static condition holds only when the explicit dependence
$p({\varrho}_{\text m})$ is linear, i.e., essentially of the form 
$p{\propto}{\varrho}_{\text m}$, since otherwise the equation of state will not 
be consistent with the given time evolution. Once a suitable equation of state 
is given, it is necessary to use the expressions (41) and (44) relating 
${\bar {\varrho}}_{\text m}$ to the passive mass density ${\varrho}_{\text m}$ and 
similarly for a relationship between ${\bar p}$ and the passive pressure $p$. 
As mentioned above, the relationship between ${\bar {\varrho}}_{\text m}$ and 
${\varrho}_{\text m}$ will be the same for photons and material particles when 
taking into account the cosmic redshift of photons.

To find how the active mass $m_t$ varies in space-time, note that we are free 
to choose the background value of the active mass far from the source to be 
$m_0$. We use this to define $G^{\rm B}$ and $G^{\rm S}$ as the constants measured
in local gravitational experiments performed far from the source at epoch 
$t_0$. Then, using the metrically static condition and equations (8) and (43), 
we find that
\eqa
m_t({\rho},t)={\bar B}^{1/2}({\rho}){\frac{t}{t_0}}m_0.
\ena
Now we can insert equations (43) and (44) into the field equation (12). Since 
${\bf {\bar K}}_t$ vanishes identically for the metrically static case [4], 
equation (13) becomes vacant and equation (14) only confirms the form (42) of 
the metric family. On the other hand, equation (12) yields 
\eqa
(1-{\frac{{\rho}^2}{{\Xi_0^2}}}){\frac{{\bar B},_{{\rho}{\rho}}}{{\bar B}}}
+{\frac{2}{\rho}}(1-{\frac{3}{2}}{\frac{{\rho}^2}{{\Xi_0^2}}})
{\frac{{\bar B},_{\rho}}{{\bar B}}}={\Big [}
{\kappa}^{\rm B}({\bar {\varrho}}^{\rm (EM)}_{\rm m}c^2+3{\bar p}^{\rm (EM)})+
{\kappa}^{\rm S}({\bar {\varrho}}^{\rm (MA)}_{\rm m}c^2+3{\bar p}^{\rm (MA)})
{\Big ]}.
\ena
From equation (21) we easily see that ${\bar A}(0)=1$. Moreover,
${\bar B},_{{\rho}{\rho}}(0)$ must be finite and we must also
have ${\bar B},_{\rho}(0)={\bar p},_{\rho}(0)=0$, yielding
${\bar A},_{\rho}(0)=0$. Furthermore, noticing that ${\rho}^{-1}{\bar B},_{\rho}$ 
must be stationary near the centre of the body, we must have
\eqa
{\bar B},_{{\rho}{\rho}}(0)={\lim}_{{\rho}{\rightarrow}0}
{\Big [}{\rho}^{-1}{\bar B},_{\rho}({\rho}){\Big ]}, 
\qquad \Rightarrow \qquad \nonumber \\
{\frac{{\bar B},_{{\rho}{\rho}}(0)}{{\bar B}(0)}}
={\frac{{\kappa}^{\rm B}}{3}}{\Big (}
{\bar {\varrho}}^{\rm (EM)}_{\rm m}(0)c^2+3{\bar p}^{\rm (EM)}(0){\Big )}
+{\frac{{\kappa}^{\rm S}}{3}}{\Big (}
{\bar {\varrho}}^{\rm (MA)}_{\rm m}(0)c^2+3{\bar p}^{\rm (MA)}(0){\Big )},
\ena
where the implication follows from equation (47). Next it is straightforward 
to show that equation (47) can be once integrated to yield
\eqa
{\bar B},_{\rho}({\rho})=
{\frac{2[G^{\rm B}{\bar M}_{t_0}^{\rm (EM)}({\rho})+
G^{\rm S}{\bar M}_{t_0}^{\rm (MA)}({\rho})]
}{c^2{\rho}^2{\sqrt{1-{\frac{{\rho}^2}{{\Xi_0^2}}}}}}}, \nonumber \\
{\bar M}_{t_0}^{\rm (MA)}({\rho}){\equiv}
{\frac{4{\pi}}{c^2}} \int_0^{\rho}{\frac{{\bar B}({\rho}')
[{\bar {\varrho}}^{\rm (MA)}_{\rm m}c^2+
3{\bar p}^{\rm (MA)}]{{\rho}'}^2d{\rho}'}
{{\sqrt{1-{\frac{{{\rho}'}^2}{{\Xi_0^2}}}}}}},
\ena
where a similar definition can be made for ${\bar M}_{t_0}^{\rm (EM)}({\rho})$. 
Integrating equation (49), using integration by parts, then yields the 
implicit solution
\eqa
{\bar B}({\rho})={\bar B}(0)-
{\frac{2[G^{\rm B}{\bar M}_{t_0}^{\rm (EM)}({\rho})+
G^{\rm S}{\bar M}_{t_0}^{\rm (MA)}({\rho})]
}{c^2{\rho}}}{\sqrt{1-{\frac{{\rho}^2}{{\Xi_0^2}}}}} \nonumber \\
+{\frac{8{\pi}}{c^4}} \int_0^{{\rho}}{\bar B}({\rho}'){\Big (}G^{\rm B}[
{\bar {\varrho}}^{\rm (EM)}_{\rm m}c^2+3{\bar p}^{\rm (EM)}]+G^{\rm S}
[{\bar {\varrho}}^{\rm (MA)}_{\rm m}c^2+3{\bar p}^{\rm (MA)}]{\Big )}
{{\rho}'}d{\rho}', \quad {\rho}{\leq}{\rho}_{\rm sf}.
\ena
To match the exterior solution (see the next section) at the surface 
${\rho}={\rho}_{\rm sf}$ of the body, we must have (setting 
$M_{t_0}^{\rm (EM)}{\equiv}{\bar M}_{t_0}^{\rm (EM)}({\rho}_{\rm sf})$ and 
$M_{t_0}^{\rm (MA)}{\equiv}{\bar M}_{t_0}^{\rm (MA)}({\rho}_{\rm sf})$)
\eqa
{\bar B}({\rho}_{\rm sf})=1-{\frac{r_{{\rm s}0}}{{\rho}_{\rm sf}}}
{\sqrt{1-{\frac{{\rho}^2_{\rm sf}}{{\Xi_0^2}}}}}, \qquad
r_{{\rm s}0}{\equiv}{\frac{2[G^{\rm B}M_{t_0}^{\rm (EM)}+
G^{\rm S}M_{t_0}^{\rm (MA)}]}{c^2}},
\ena
(where $r_{{\rm s}0}$ is the generalized Schwarzschild radius at epoch $t_0$
of the body), so that equation (50) yields the ``normalizing'' condition
\eqa
{\bar B}(0)+{\frac{8{\pi}}{c^4}} 
\int_0^{{\rho}_{\rm sf}}{\bar B}({\rho}'){\Big (}G^{\rm B}[
{\bar {\varrho}}^{\rm (EM)}_{\rm m}c^2+3{\bar p}^{\rm (EM)}]+G^{\rm S}
[{\bar {\varrho}}^{\rm (MA)}_{\rm m}c^2+3{\bar p}^{\rm (MA)}]{\Big )}
{{\rho}'}d{\rho}'=1.
\ena
Inserting the condition (52) into equation (50) then yields the final form of
the implicit solution, i.e.,
\eqa
{\bar B}({\rho})=1-
{\frac{2[G^{\rm B}{\bar M}_{t_0}^{\rm (EM)}({\rho})+
G^{\rm S}{\bar M}_{t_0}^{\rm (MA)}({\rho})]
}{c^2{\rho}}}{\sqrt{1-{\frac{{\rho}^2}{{\Xi_0^2}}}}} \nonumber \\
-{\frac{8{\pi}}{c^4}} \int_{\rho}^{{\rho}_{\rm sf}}
{\bar B}({\rho}'){\Big (}G^{\rm B}[
{\bar {\varrho}}^{\rm (EM)}_{\rm m}c^2+3{\bar p}^{\rm (EM)}]+G^{\rm S}
[{\bar {\varrho}}^{\rm (MA)}_{\rm m}c^2+3{\bar p}^{\rm (MA)}]{\Big )}
{{\rho}'}d{\rho}', \quad {\rho}{\leq}{\rho}_{\rm sf}.
\ena
The implicit solution (53) is not very useful, so one should rather try to
solve equations (45) and (47) numerically for ${\bar p}({\rho})$ and
${\bar B}({\rho})$ using equation (49). One may proceed as follows. Given a
suitable equation of state, specify the boundary conditions 
${\bar p}^{\rm (EM)}(0)$ and ${\bar p}^{\rm (MA)}(0)$ at the centre of the body for
some arbitrary time in addition to some value for ${\bar B}(0)$ chosen
as an initial value for iteration. Now integrate equation (45) outwards from 
${\rho}=0$ using equation (49) until the pressure vanishes. The surface of the 
body ${\rho}_{\rm sf}$ is then reached. But the condition (52) must be 
satisfied. If it is not, add a constant to ${\bar B}({\rho})$ everywhere such 
that equation (52) holds. Then repeat the calculations with the new value of 
${\bar B}(0)$. Iterate until a value for ${\bar B}(0)$ is found that satifies 
equation (52) (to desired accuracy) without any further adjustments. Once we 
have finished these calculations for an arbitrary time, we know the time 
evolution of the system from equation (45) and we have found the family 
${\bf {\bar g}}_t$ numerically inside the body. To find the corresponding 
family ${\bf g}_t$ one uses the method described in [3, 4].

As already mentioned; to have a metrically static system, it is necessary to 
specify an equation of state of the type $p{\propto}{\varrho}_{\text m}$ 
(potential implicit dependences not included) since this ensures that 
${\bar {\varrho}}_{\text m}$ and ${\bar p}$ are independent of $t$. This implies 
that a spherical gravitationally bound body made of perfect fluid obeying an 
equation of state of type $p{\propto}{\varrho}_{\text m}$ will expand according 
to the Hubble law. But for bodies made of perfect fluid obeying other equations 
of state (degenerate star matter for example), the expansion may induce 
instabilities; mass currents will be set up and such systems cannot be 
metrically static. However, for the metrically static case, the gravitational 
field interior to the body will expand along with the fluid; this is similar 
to the expansion of the exterior gravitational field found in the next section.
Since the equation of state for an ideal gas has the required form and since 
non-degenerate star matter is reasonably well approximated by an ideal gas, it 
should not be too unrealistic to apply the metrically static condition 
to main-sequence stars. (The metrically static condition holds in general for
an ideal gas even if the gas is not isothermal; i.e., even if the temperature
depends on the pressure so that the equation of state takes a polytropic 
form.) 

We finish this section by estimating how the cosmic expansion will affect a 
spherically symmetric body made of perfect fluid obeying an equation of state 
of the form $p{\propto}{\varrho}_{\text m}^{\gamma}$ (e.g., a polytrope made of
degenerate matter). To do that, we assume that the fluid-dynamical effects on 
the gravitational field coming from instabilities can be neglected; i.e., we 
assume that the body can be treated as being approximately in hydrostatic 
equilibrium for each epoch $t$. (Effects coming from gravitational heating of 
the body due to contraction are also neglected.) To justify this approximation 
we work in the Newtonian limit, so that we can use the approximations
${\rho}{\approx}r$, ${\bar {\varrho}}_{\text m}{\approx}{\varrho}_{\rm m}$, 
${\bar p}{\approx}p$ and $G^{\rm B}{\bar {\varrho}}^{\rm (EM)}_{\text m}+
G^{\rm S}{\bar {\varrho}}^{\rm (MA)}_{\text m}{\approx}G_{\rm N}{\varrho}_{\text m}$, 
where $G_{\rm N}$ is Newton's constant. Applying these approximations, we take 
the Newtonian limits of equations (45) and (47), getting
\eqa
{\frac{d}{dr}}{\frac{r^2}{{\varrho}_{\rm m}}}
{\frac{dp}{dr}}=-4{\pi}G_{\rm N}r^2{\varrho}_{\rm m0},
\ena
where ${\varrho}_{\rm m0}$ is the density field ${\varrho}_{\rm m}$ at the 
present epoch $t_0$. As it stands, equation (54) is valid only for epoch $t_0$ 
if ${\gamma}{\neq}1$. However, by making the substitution $r{\rightarrow}
({\frac{t_0}{t}})^{\frac{1}{3{\gamma}-4}}r{\equiv}{\bar r}$, 
${\varrho}_{\rm m0}{\rightarrow}({\frac{t}{t_0}})^{\frac{3}{3{\gamma}-4}}
{\varrho}_{\rm m0}{\equiv}{\varrho}_{\rm m}$, 
$G_{\rm N}{\rightarrow}{\frac{t}{t_0}}G_{\rm N}{\equiv}G_t$, equation (54) may 
effectively be transformed from epoch $t_0$ to epoch $t$ and it may then be 
applied to the body at each fixed epoch $t$ even if ${\gamma}{\neq}1$. 
Equation (54) then becomes equivalent to its counterpart in Newtonian theory 
except for a variable $G_t$. Thus the usual Newtonian analysis of polytropes 
[7] applies, but with $G_{\rm N}$ variable. And as consequences of this we see 
that the physical radius of a polytrope will actually {\em shrink} with epoch 
(if ${\gamma}>{\frac{4}{3}}$), and that the Chandrasekhar mass limit will 
decrease with epoch. Thus any white dwarf made of degenerate matter is 
predicted to shrink with epoch and eventually explode as a type Ia supernova 
when the Chandrasekhar mass gets close to the mass of the white dwarf. In 
particular this should happen to isolated white dwarfs, so according to 
quasi-metric theory it is not necessary to invoke mass accretion from exterior 
sources to ignite type Ia supernovae.
\subsection{The exterior field}
To find the function ${\bar B}({\rho})$ for the exterior field, 
we must solve equation (47) without sources, i.e.,
\eqa
(1-{\frac{{\rho}^2}{{\Xi_0^2}}}){\frac{{\bar B},_{{\rho}{\rho}}}{{\bar B}}}
+{\frac{2}{\rho}}(1-{\frac{3}{2}}{\frac{{\rho}^2}{{\Xi_0^2}}})
{\frac{{\bar B},_{\rho}}{{\bar B}}}=0.
\ena
However, before we try to solve equation (55), it is important to notice that 
no solution of it can exist on a whole FHS (except for the trivial solution 
${\bar B}=$constant), according to the maximum principle applied to a closed 
Riemannian 3-manifold. The reason for this is the particular form of equation 
(55), see reference [4] and references therein for justification. This means 
that {\em in quasi-metric theory, isolated systems cannot exist except as an 
approximation.} But even if a non-trivial solution of equation (55) does not 
exist on a whole FHS, we may try to find a solution valid in some finite 
region of a FHS. That is, we want to find a solution of equation (55) in the 
region ${\rho}_{\rm sf}{\leq}{\rho}{\leq}{\Xi}_0$, with the chosen boundary 
condition ${\bar B}({\Xi}_0)=1$ to have a correspondence with the limiting case
where the mass of the central source goes to zero. The limited region of 
validity of such a solution is not of much concern since the approximation made
by assuming an isolated system is physically reliable only if 
${\frac{{\rho}}{{\Xi}_0}}{\ll}1$.

Since equation (55) is linear, it is easy to solve and the unique solution,
given said boundary condition, is
\eqa
{\bar B}({\rho})=1-
{\frac{r_{{\rm s}0}}{{\rho}}}{\sqrt{1-{\frac{{\rho}^2}{{\Xi_0^2}}}}} \qquad
{\rho}_{\rm sf}{\leq}{\rho}{\leq}{\Xi}_0.
\ena
The solution (56) will be more useful if we rather express ${\bar B}$ as a 
function of $r={\sqrt{\bar B}}{\rho}$, since that radial coordinate was used 
in section 2.2. We then find that
\eqa
{\bar B}(r)={\Big (}{\sqrt{1+({\frac{r_{\text s0}}{2r}})^2-
{\frac{r^2}{{\Xi}_0^2}}}}-{\frac{r_{\text s0}}{2r}}{\Big )}^2+
{\frac{r^2}{{\Xi}_0^2}}, \qquad 
r_{\rm sf}{\equiv}{\sqrt{{\bar B}({\rho}_{{\rm sf}})}}{\rho}_{\rm sf}
{\leq}r{\leq}{\Xi}_0.
\ena
Moreover, from equations (57) and (21) we find that
\eqa
{\bar A}(r)={\Big [}1+({\frac{r_{\text s0}}{2r}})^2-{\frac{r^2}{{\Xi}_0^2}}
{\Big ]}^{-1}{\bar B}(r).
\ena
For small $r$, we may write expressions (57) and (58) as series expansions, 
i.e., as perturbations around the analogous problem in a Minkowski background. 
But in contrast to the analogous case with a Minkowski background, there 
exists the extra scale ${\Xi}_0$ in addition to the generalized Schwarzschild 
radius $r_{\text s0}$ defined in equation (51). To begin with, we try to model 
the gravitational field exterior to galactic-sized objects, so we may assume 
that the typical scales involved are determined by 
${\frac{r}{{\Xi}_0}}\raisebox{-2pt}{\,$\stackrel{>}{\sim}$\,}
{\frac{r_{\text s0}}{r}}$; this criterion tells how to compare the importance 
of the different terms of the series expansion. One may straightforwardly show 
that series expansions of equations (57) and (58) yield
\eqa
{\bar B}(r)&=&1-{\frac{r_{\text s0}}{r}}+{\frac{r_{\text s0}^2}{2r^2}}+
{\frac{r_{\text s0}r}{2{\Xi}_0^2}}-{\frac{r_{\text s0}^3}{8r^3}}+{\cdots},
{\qquad} {\Rightarrow} {\nonumber} \\
{\bar A}(r)&=&1-{\frac{r_{\text s0}}{r}}+
{\frac{r_{\text s0}^2}{4r^2}}+{\frac{r^2}{{\Xi}_0^2}}+{\cdots}.
\ena
To construct the family ${\bf g}_t$ as described in [3], we need the quantity
$v(r)$, which for spherically symmetric systems takes the form [4]
\eqa
v(r)={\bar y}_tr{\sqrt{{\bar h}_{(t)rr}}}={\frac{cr}{2}}
{\frac{{\bar B}'(r)}{{\bar B}(r)}}=
{\frac{r_{\text s0}}{2r}}{\frac{c}{{\sqrt{1+({\frac{r_{\text s0}}{2r}})^2-
{\frac{r^2}{{\Xi}_0^2}}}}}}=
{\frac{r_{\text s0}c}{2r}}[1+O({\frac{r_{\text s0}^2}{r^2}})].
\ena
We notice that $v(r)$ does not depend on $t$. The functions $A(r)$ and $B(r)$ 
are found from the relations (valid for spherically symmetric systems [4])
\eqa
A(r)={\Big(}{\frac{1+{\frac{v(r)}{c}}}{1-{\frac{v(r)}{c}}}}{\Big)}^2
{\bar A}(r), {\qquad}
B(r)={\Big(}1-{\frac{v^2(r)}{c^2}}{\Big)}^2{\bar B}(r).
\ena
From equations (57), (58) and (61) we then get
\eqa
B(r)={\frac{(1-{\frac{r^2}{{\Xi}_0^2}})^2}{(1+({\frac{r_{\text s0}}{2r}})^2-
{\frac{r^2}{{\Xi}_0^2}})^2}}{\Big [}
{\Big (}{\sqrt{1+({\frac{r_{\text s0}}{2r}})^2-
{\frac{r^2}{{\Xi}_0^2}}}}-{\frac{r_{\text s0}}{2r}}{\Big )}^2+
{\frac{r^2}{{\Xi}_0^2}}{\Big ]},
\ena
\eqa
A(r)={\frac{{\Big (}{\sqrt{1+({\frac{r_{\text s0}}{2r}})^2-
{\frac{r^2}{{\Xi}_0^2}}}}+{\frac{r_{\text s0}}{2r}}{\Big )}^2}{1+
({\frac{r_{\text s0}}{2r}})^2-{\frac{r^2}{{\Xi}_0^2}}}}{\Big {\{ }}1+
{\frac{{\Big (}{\sqrt{1+({\frac{r_{\text s0}}{2r}})^2-
{\frac{r^2}{{\Xi}_0^2}}}}+{\frac{r_{\text s0}}{2r}}{\Big )}^2}{{\Big (}1-
{\frac{r^2}{{\Xi}_0^2}}{\Big )}^2}}{\frac{r^2}{{\Xi}_0^2}}
{\Big {\}}}.
\ena
Note that, although $B(r)$ increases for small $r$, for large $r$ it 
eventually reaches a maximum value and then decreases towards zero when 
$r{\rightarrow}{\Xi}_0$. This is merely a curious effect due to the global 
curvature of space and the unrealistic assumption that an isolated source
determines the gravitational field at cosmological distances. That is, it is 
utterly unrealistic to assume that an isolated source dominates the 
gravitational field over cosmological scales and that this source has been 
present since the beginning of time. Thus the from equation (62) inferred 
gravitational repulsion on cosmological scales is nothing but an unrealistic 
model artefact.

It is useful to have series expansions for $B(r)$ and $A(r)$. Putting
these into a family of line elements we find
\eqa
ds^2_t&=&-{\bigg(}1-{\frac{r_{\text s0}}{r}}
+{\frac{r_{\text s0}r}{2{\Xi}_0^2}}+{\frac{3r_{\text s0}^3}{8r^3}}+{\cdots}
{\bigg)}(dx^0)^2 \nonumber \\
&&+({\frac{t}{t_0}})^2{\bigg(}{\Big\{}1+{\frac{r_{\text s0}}{r}}+
{\frac{r^2}{{\Xi}_0^2}} + {\frac{r^2_{\text s0}}{4r^2}}+{\cdots}
{\Big\}}dr^2+r^2d{\Omega}^2{\bigg)}.
\ena
This expression represents the wanted metric family as a series expansion. 
Note in particular the fact that all spatial dimensions expand whereas the 
corresponding Newtonian potential $-U=-{\frac{c^2r_{\text s0}}{2r}}$ (to
Newtonian order) remains constant for a fixed FO. This means that the physical 
radius of any circle orbit (i.e., with $r$ constant) increases but such that 
the orbital speed remains constant. That is, {\em the (active) mass of the 
central object as measured by distant orbiters increases to exactly balance 
the effect on circle orbit velocities of expanding circle radii.} This is not 
as outrageous as it may seem due to the extra formal variation of atomic units
built into our theory. So this result is merely a consequence of the fact that
the coupling between matter and geometry depends directly on the formal 
variation via the field equations.

What is measured by means of distant orbiters is not the ``bare'' mass 
$M^{\rm (MA)}_t+M^{\rm (EM)}_t$ itself but rather the combination 
$M^{\rm (MA)}_tG^{\rm S}+M^{\rm (EM)}_tG^{\rm B}$. We have, however, {\em defined} 
$G^{\rm S}$ and $G^{\rm B}$ to be constants. And as might be expected, it turns 
out that the variation of $M^{\rm (MA)}_tG^{\rm S}+M^{\rm (EM)}_tG^{\rm B}$ with $t$ 
as inferred from equation (64) is exactly that found directly from the formal 
variation of the active masses $M_t^{\rm (MA)}={\frac{t}{t_0}}M_{t_0}^{\rm (MA)}$ 
and $M_t^{\rm (EM)}={\frac{t}{t_0}}M_{t_0}^{\rm (EM)}$ with $t$ by 
using equation (8). This means that the dynamically measured mass 
increase should not be taken as an indication of actual particle creation but 
that the general dynamically measured mass scale should be taken to change via 
a linear increase of $M_t^{\rm (MA)}$ and $M_t^{\rm (EM)}$ with $t$, and that this 
is directly reflected in the gravitational field of the source. That is, 
measured in atomic units, active mass increases linearly with epoch in 
accordance with equation (8) (for photon energy this only works when including
the cosmic redshift of photon energy in an expanding source).

Any dynamical measurement of the mass of a central object by means of 
distant orbiters does not represent a local test experiment. Nevertheless, the 
dynamically measured mass increase thus found is just as ``real" as the 
expansion in the sense that neither should be neglected on extended scales.
This must be so since in quasi-metric relativity, the global scale increase 
and the dynamically measured mass increase are two different aspects of the 
same basic phenomenon.
\section{The effects of cosmic expansion on gravitation} 
\subsection{Shapes of orbits and rotational curves}
We now explore which kinds of free-fall orbits we get from equation (64) 
and the equations of motion. To begin with, we find the shape of the rotational
curve as defined from the 3-velocities ${\bf w}_t$ of the circle orbits. (The 
4-velocities ${\bf u}_t$ may be split up into pieces respectively orthogonal 
to and intrinsic to the FHSs according to the formula
${\sqrt{1-{\frac{w^2}{c^2}}}}{\bf u}_t=c{\bf n}_t+{\bf w}_t$.) Since equation 
(36) has no time dependence for such orbits we can do a standard calculation 
[7], and the result is that orbital speed $w$ varies as
\eqa
w(r)={\frac{t}{t_0}}r{\frac{d{\phi}}{dt}}
{\sqrt{1-{\frac{w^2}{c^2}}}}{\frac{dt}{d{\tau}_t}}=
B^{-1/2}(r){\frac{t}{t_0}}r{\frac{d{\phi}}{dt}}
={\sqrt{\frac{B'(r)r}{2B(r)}}}c,
\ena
where the second step follows from the formula ${\frac{d{\tau}_t}{dt}}=
{\sqrt{B(r)-{\frac{t^2}{t^2_0}}r^2({\frac{d{\phi}}{cdt}})^2}}$ (obtained from 
equation (20) for circular motion) together with a consistency requirement.
However, when we apply equation (65) to the metric family (64), we get a result 
essentially identical to the standard Keplerian rotational curve; the only 
effect of the dynamically measured mass increase and the non-kinematical 
expansion is to increase the scale but such that the shape of the rotational 
curve remains unaffected. It is true that $B(r)$ as found from equation (64) 
contains a term linear in $r$ in addition to terms falling off with increasing
$r$; in reference [8] it is shown that such a linear term may be successfully 
used to model the asymptotically non-Keplerian rotational curves of spiral 
galaxies. But the numerical value of the linear term found from equation (64) 
is too small by a factor of order $10^{-10}$ to be able to match the data. So 
at least the simple model considered in this paper is unable to explain the 
asymptotically non-Keplerian rotational curves of spiral galaxies from first 
principles. 

Another matter is how the time dependence in the equations of motion will
affect the time histories and shapes of more general orbits than the circle 
orbits. Clearly, time histories will be affected as can be seen directly from
equation (36). However, to see if this is valid for shapes as well, we may 
insert equation (35) into equation (36) to obtain $r$ as a function of 
${\phi}$. This yields the equation
\eqa
{\frac{A(r)}{r^4}}{\Big(}{\frac{dr}{d{\phi}}}{\Big)}^2+{\frac{1}{r^2}}-
{\frac{1}{J^2B(r)}}=-{\frac{E}{J^2}},
\ena
and this is identical to the equation valid for the case of a single 
spherically symmetric static metric [7]. Thus the shapes of free-fall orbits
are unaffected by the global non-kinematical expansion present in the metric 
family (64).
\subsection{Expanding space and the solar system}
One may try to apply the metric family (64) to the solar system by using it 
to describe the gravitational field of the Sun (when gravitational fields of 
other solar system bodies than the Sun are neglected). That is, as a good 
approximation, we may neglect the gravitational effects of the galaxy and 
treat the solar system as an isolated system. But the solar system is not at 
rest with respect to the cosmic rest frame; this follows from the observed 
dipole in the cosmic microwave background radiation. However, as long as the 
solar system can be treated as approximately isolated, its velocity with 
respect to the cosmic rest frame is not crucial when solving the field 
equations (see the discussion at the end of section 2.1). So as a good 
approximation, we may neglect the solar system's motion with respect to the 
cosmic rest frame and use the metric family (64) to describe the gravitational 
field of the Sun. Also, since the solar system is small, we can neglect any 
dependence on ${\Xi}_0$ (and thus the global curvature of space). The errors 
made by neglecting terms depending on ${\Xi}_0$ in equation (64) are 
insignificant since the typical scales involved for the solar system are 
determined by ${\frac{r}{{\Xi}_0}}\raisebox{-2pt}{\,$\stackrel{<}{\sim}$\,}
{\frac{r^3_{\text s0}}{r^3}}$. Equation (64) then takes the form
\eqa
ds^2_t=-{\Big(}1-{\frac{r_{\text s0}}{r}}+
O({\frac{r^3_{\text s0}}{r^3}}){\Big)}(dx^0)^2
+({\frac{t}{t_0}})^2{\Big(}{\{}1+{\frac{r_{\text s0}}{r}}+
O({\frac{r^2_{\text s0}}{r^2}}){\}}dr^2+r^2d{\Omega}^2{\Big)}.
\ena
Equation (66) shows that the shapes of orbits are unaffected by the expansion; 
this means that all the classical solar system tests come in just as for the 
analogous case of a Minkowski background [4]. However, we get at least one 
extra prediction (irrespective of whether or not the galactic gravitational
field can be neglected); from equation (67) we see that the effective distance 
between the Sun and any planet is predicted to have been smaller in the past. 
That is, the spatial coordinates are co-moving rather than static, thus average
distances (measured in atomic units) between bodies within the solar system 
are predicted to show a secular increase as a consequence of the cosmic 
expansion. For example, the distance between the Sun and the Earth at the time 
of its formation may have been almost 50{\%} smaller than today. But since 
main-sequence stars are predicted to expand according to quasi-metric theory, a 
small Earth-Sun distance should not be incompatible with paleo-climatic data, 
since the Sun is expected to have been smaller and thus dimmer in the past. 
Actually, since neither the temperature at the centre of the Sun (as estimated
from the virial theorem), nor the radiation energy gradient times the mean 
free path length of a photon depend on $t$, the cosmic luminosity evolution of
the Sun should be determined from the cosmic expansion of its surface area as
long as the ideal gas approximation is sufficient. And this luminosity 
evolution exactly balances the effects of an increasing Earth-Sun distance on 
the effective solar radiation received at the Earth.

However, an obvious question is if the predicted effect of the expansion on 
the time histories of non-relativistic orbits is compatible with the observed 
motions of the planets. In order to try to answer this question, it is 
illustrating to calculate how the orbit period of any planet depends on $t$. 
For simplicity, consider a circular orbit $r=R=$constant. Equation (35) then 
yields
\eqa
{\frac{d{\phi}}{dt}}={\frac{t_0}{t}}B(R)R^{-2}Jc.
\ena
Now integrate equation (68) one orbit period $T<<t$ (i.e., from $t$ to $t+T$). 
The result is
\eqa
T(t)=t({\exp}{\Big [}{\frac{T_{\text{GR}}}{t_0}}{\Big ]}-1)=
{\frac{t}{t_0}}T_{\text{GR}}(1+{\frac{T_{\text{GR}}}{2t_0}}+{\cdots}), 
{\qquad} T_{\text{GR}}{\equiv}{\frac{2{\pi}R^2}{cJB(R)}},
\ena
where $T_{\text{GR}}$ is the orbit period as predicted from General Relativity.
From equation (69), we see that (sidereal) orbit periods are predicted to 
increase linearly with cosmic scale, i.e.,
\eqa
T(t)={\frac{t}{t_0}}T(t_0), {\qquad} {\frac{dT}{dt}}={\frac{T(t_0)}{t_0}},
\ena
and such that any ratio between periods of different orbits remains constant.
In particular, equation (70) predicts that the (sidereal) year $T_{\text E}$ 
should be increasing with about 2.5 ms/yr and the martian year $T_{\text M}$ 
should be increasing by about 4.7 ms per martian year at the present epoch. 
This should be consistent with observations since the observed difference in 
the synodical periods of Mars and the Earth is accurate to about 5 ms. 

To compare predictions coming from equation (67) against timekeeping data, one 
must also take into account the predicted cosmological contribution to the 
spin-down of the Earth. If one assumes that the gravitational source of the 
exterior field (67) is stable with respect to internal collapse (as for a
source made of ideal gas), i.e., that possible instabilities generated by the
expansion can be neglected, one may model this source as a uniformly expanding 
sphere. Due to the increase with time of active mass, the angular momenta of 
test particles moving in the exterior field (67) increase linearly with cosmic
scale. This also applies to the angular momentum $L_{\text s}$ of a spinning
source made of ideal gas [9], that is
\eqa
L_{\text s}(t)={\frac{t}{t_0}}L_{\text s}(t_0), {\qquad} 
{\frac{dL_{\text s}}{dt}}={\frac{1}{t}}L_{\text s}=(1+O(2))HL_{\text s},
\ena
where the term $O(2)$ is of post-Newtonian order and where the locally 
measured Hubble parameter $H$ is defined by $H{\equiv}{\frac{1}{Nt}}$, or 
equivalently (${\tau}_{\cal F}$ is the proper time of the local FO)
\eqa
H{\equiv}{\frac{t_0}{t}}{\frac{d}{d{\tau}_{\cal F}}}({\frac{t}{t_0}})
={\frac{ct_0}{t}}{\Big (}{\sqrt{B(r)}}{\Big )}^{-1}
{\frac{d}{dx^0}}({\frac{t}{t_0}})={\Big (}{\sqrt{B(r)}}t{\Big )}^{-1}.
\ena
Since the moment of inertia $I{\propto}MR_{\text s}^2$, where $M$ is the 
passive mass and $R_{\text s}$ is the measured radius of the sphere, we must 
have (neglecting terms of post-Newtonian order)
\eqa
{\frac{dR_{\text s}}{dt}}=HR_{\text s}, {\qquad} 
{\frac{d{\omega}_{\text s}}{dt}}=-H{\omega}_{\text s}, {\qquad} 
{\frac{dT_{\text s}}{dt}}=HT_{\text s},
\ena
where ${\omega}_{\text s}$ is the spin circle frequency and $T_{\text s}$ is 
the spin period of the sphere. (To show  equation (73), use the definition 
$L_{\text s}=I{\omega}_{\text s}$.) This means that the spin period of a
sphere made of ideal gas increases linearly with $t$ due to the cosmic 
expansion. Does this apply to the Earth as well? The Earth is not made of
ideal gas, so the cosmic expansion may induce instabilities, affecting its
(sidereal) spin period $T_{\rm sE}$. However, here we assume that the Earth's
mantle is made of a material which may be approximately modelled as a perfect
fluid obeying an equation of state close to linear. Then, if this assumption 
holds, the Earth should be expanding close to the Hubble rate according to the 
discussion following equation (54). Moreover, averaged over long time spans, 
shorter timescale effects of instabilities on $T_{\rm sE}$ should average out
to a good approximation. We may also assume that there is no significant tidal 
friction since given the cosmic contribution, this would be inconsistent with 
the observed so-called mean acceleration ${\dot n}_{\rm m}$ of the Moon (see 
below). We then get
\eqa
{\frac{dT_{\rm sE}}{dt}}=HT_{\rm sE}, \qquad \Rightarrow \qquad
T_{\rm sE}(t)={\frac{t}{t_0}}T_{\rm sE}(t_0).
\ena
From equation (74), we may estimate a cosmic spin-down of the Earth at the 
present epoch to be about 0.68 ms/cy (using $H{\sim}2.5{\times}
10^{-18}$ s$^{-1}$). To see if this is consistent with the assumption that
the dominant contribution is due to cosmic effects, we may compare to results 
obtained from historical observations of eclipses from AD 1000 and onwards. 
These observations can be used to infer a lengthening of the day of about 
$1.4$ ms/cy [10], whereas an average over the last 2700 years shows a value of 
about $1.70$ ms/cy [11]. But the interpretation of the historical data depends 
on an assumed vale of $-26^{{\prime}{\prime}}$/cy$^2$ for the tidal contribution 
${\dot n}_{\rm tid}$ to the mean acceleration ${\dot n}_{\rm m}$ of the Moon 
(moreover, other significant (theory-dependent) contributions to 
${\dot n}_{\rm m}$ are neglected without justification, see below). This value 
of ${\dot n}_{\rm tid}$ corresponds to a {\em calculated} lengthening of the day 
(using standard theory) of about $2.3$ ms/cy [11]; thus the agreement with the 
values inferred from the historical data is not very good without invoking a 
secular {\em shortening} of the length of the day of non-tidal origin. On the 
other hand, the QMF yields a value of about $-13.6^{{\prime}{\prime}}$/cy$^2$ for 
${\dot n}_{\rm m}$ (see below). Reinterpreting the historical data using this 
value, yields a correction to the lenghtening of the day of about $-0.62$ 
ms/cy, i.e., the observations could indicate a lengthening of the day of about 
$0.78$ ms/cy and $1.08$ ms/cy, respectively, rather than the values given 
above. This means that the values obtained from the historical observations are
theory-dependent and that the secular spin-down of the Earth may be only about 
half of the currently accepted value. Note that such a theory-dependence also
affects the comparison of equation (74) to results obtained from sedimentary 
tidal rhythmities [12].

Another quantity that can be calculated from equation (74) is the number of 
days $N_{\text y}$ in one (sidereal) year $T_E$. This is found to be constant
since
\eqa
T_{\text E}=N_{\text y}T_{\rm sE}={\frac{t}{t_0}}T_{\text E}(t_0), {\qquad} 
\Rightarrow \qquad {\frac{dN_{\text y}}{dt}}=0. 
\ena
Moreover, the number of the days $N_{\text m}$ in one (sidereal) month 
$T_{\text m}$ can be calculated similarly. We then get
\eqa
T_{\text m}=N_{\text m}T_{\rm sE}={\frac{t}{t_0}}T_{\text m}(t_0), {\qquad} 
\Rightarrow \qquad
{\frac{dN_{\text m}}{dt}}=0.
\ena
That $N_{\text y}$ and $N_{\text m}$ are predicted to be constant is not in 
agreement with standard (model-dependent) interpretations of
paleo-geological data [6, 12]. (The predicted constancy of the ratio 
$N_{\text y}/N_{\text m}$ agrees well with a standard interpretation of the 
data, though.) In addition to the assumption that active masses do not vary
with time, the assumption that $T_{\text E}$ is constant is routinely used in 
the interpretation of tidal rhythmities and fossil coral growth data; in 
particular this applies to [12], where one has explicitly used this assumption 
when calculating $N_{\text y}$ from the data. However, values determined 
directly from the rhythmite record presented in [12] and the predictions given 
here, usually agree within two standard deviations when the predicted variable 
length of the year is taken into account.

As mentioned above, the mean acceleration of the Moon, 
${\dot n}_{\text m}{\equiv}{\frac{d}{dt}}n_{\text m}$ (where $n_{\text m}$ 
is the mean geocentric angular velocity of the Moon as observed from its 
motion) is a very important quantity for calculating the evolution of the
Earth-Moon system. From equation (76) we find the quasi-metric prediction
\eqa
n_{\text m}(t)={\frac{d{\phi}_{\text m}}{dt}}=
{\frac{t_0}{t}}n_{\text m}(t_0), \qquad \Rightarrow \qquad
{\dot n}_{\text m}{\equiv}{\frac{d}{dt}}n_{\text m}=-Hn_{\text m},
\ena
and inserting the observed value 0.549$^{{\prime}{\prime}}$/s for 
$n_{\text m}$ at the present epoch, we get the corresponding cosmological 
contribution to ${\dot n}_{\text m}$, namely about 
$-$13.6$^{{\prime}{\prime}}$/cy$^2$. This value may be compared to the value 
$-$13.74$^{{\prime}{\prime}}$/cy$^2$ obtained from fitting LLR data to a model 
based on the lunar theory ELP [13]. Note that this second value is the 
{\em total} mean acceleration, wherein other modelled (positive) contributions 
are included. These other contributions amount to about 
12.12$^{{\prime}{\prime}}$/cy$^2$ and are mainly attributed to the secular 
variation of the solar eccentricity due to (indirect) planetary perturbations 
[13]. When these contributions are removed, one deduces a {\em tidal} 
contribution ${\dot n}_{\rm tid}$ of about $-$25.86$^{{\prime}{\prime}}$/cy$^2$ to 
${\dot n}_{\text m}$ [13]. Similar values for ${\dot n}_{\rm tid}$ as inferred 
from LLR data have been found in, e.g., [14] ($-$25.9$^{{\prime}{\prime}}$/cy$^2$).
We see that in absolute values, the quasi-metric result is smaller than the 
tidal term inferred from LLR data using standard theory. But the non-tidal 
secular contributions to ${\dot n}_{\rm m}$ can be treated as model-dependent
since they are calculated from the ELP theory based on an inferred secular
acceleration of the mean longitude of the Earth's perihelion ${\dot n}_{\rm pE}$
of about $1.06''$/cy$^2$ [13]. However, said inferred value is not valid in 
quasi-metric theory since this value most likely follows from mismodelling the 
observational data; an alternative explanation of the data can be 
found from the predicted cosmic expansion of the Earth's orbit (see below). 
So it is in principle possible to omit both tidal and the traditional 
non-tidal secular contributions to ${\dot n}_{\rm m}$ and construct a 
quasi-metric model containing only the cosmic contribution. And as shown above,
such a model fits the data well.

We may also use Hubble's law directly to calculate the secular recession 
${\dot a}_{\rm qmr}$ of the Moon due to the global cosmic expansion; this 
yields about 3.0 cm/yr whereas the value ${\dot a}_{\rm tid}$ inferred from LLR 
using standard theory is (3.82${\pm}$0.07) cm/yr [14]. To see if the difference
between these results can be easily explained in terms of model-dependence, we 
notice that in standard theory, ${\dot n}_{\rm tid}$ represents the value 
${\dot n}_{\text m}$ would have had if the Earth-Moon system were isolated. 
Therefore ${\dot n}_{\rm tid}$ enters into an expression found by taking the 
time derivative of Kepler's third law. On the other hand, the quasi-metric 
model includes the cosmic contribution ${\dot n}_{\rm qmr}$ only, so that 
quantity enters into a similar expression. Taking into account the fact that 
active masses increase linearly with time according to quasi-metric theory, we 
find the relationship
\eqa
{\dot a}_{\rm tid}={\frac{2}{3}}{\frac{{\dot n}_{\rm tid}}
{{\dot n}_{\rm qmr}}}{\dot a}_{\rm qmr},
\ena
which is quite consistent with the numerical values given above. It thus seems
that there is a simple explanation of the fact that ${\dot a}_{\rm tid}$ as 
inferred from LLR data using standard theory differs from ${\dot a}_{\rm qmr}$ 
as found from Hubble's law. In other words, analysing the LLR data within the
QMF yields, to within one standard deviation, that the recession of the 
Moon follows Hubble's law.

Note once more that, whereas the secular recession of the Moon and its mean 
acceleration have traditional explanations based on tidal friction, these 
explanations are not confirmed by direct evidence. That is, tidal friction is 
of nature a mesoscopic phenomenon and it should in principle be possible to 
measure the tidal energy dissipated in the Earth's oceans. But since no 
mesoscopic measurements confirming the tidal friction scenario exist so far
[15], there are no restrictions on interpreting the secular evolution of the 
Earth-Moon system as due to cosmological effects.

The apparent constancy of the sidereal year (as indicated by astronomical 
observations of the Sun and Mercury since about AD 1680) represents the
observational basis for adopting the notion that ephemeris time (i.e., the 
time scale obtained from the observed motion of the Sun) is equal to atomic 
time (plus a conventional constant), but different from so-called universal 
time (any time scale based on the rotation of the Earth). But from 
equation (70) we see that ephemeris time should be scaled with a factor 
${\frac{t}{t_0}}$ compared to atomic time according to quasi-metric 
relativity. From equation (74) we see that averaged over long time 
spans, universal time should also be scaled with a factor ${\frac{t}{t_0}}$
compared to atomic time, as should any conventional constant difference 
between ephemeris time and universal time. Thus the predicted effect of the 
spin-down of the Earth and the expansion of the Earth's orbit is a seemingly 
inconsistency between gravitationally measured time and time measured by an 
atomic clock. But within the Newtonian framework, any secular changes in the 
Earth-Moon system are explained in terms of tidal friction (and external 
perturbations), so seemingly secular inconsistencies between different time 
scales may be blamed on the variable rotation of the Earth. In practice this 
means introducing leap seconds. Given the fact that leap seconds are routinely 
used to adjust the length of the year, the predicted differences between 
gravitational time and atomic time should be consistent with observations. In 
particular, the extra time corresponding to an increasing year as predicted 
from the quasi-metric model may easily be hidden into the declining number of 
days in a year as predicted from standard theory.

The prediction that the year increases with about $2.5$ ms at the present epoch
due to the expansion of the Earth's orbit, corresponds to a heliocentric mean 
angular acceleration ${\dot n}_{\rm E}$ of about $-1.0''$/cy$^2$. At first 
glance, this seems wildly inconsistent with the orbital motion of the 
Earth-Moon barycentre as inferred from the ELP theory and LLR data since the 
published value of ${\dot n}_{\rm E}$ is only about $-0.040''$/cy$^2$ [13]. On 
the other hand, a value for ${\dot n}_{\rm pE}$ of about $1.06''$/cy$^2$ is also 
inferred from the ELP theory and the LLR data [13]. But since this value is 
fitted and thus model-dependent, in addition to having the right size and sign,
it is very possible that it has been used to hide a large (negative) value of 
${\dot n}_{\rm E}$ of about $-1.0''$/cy$^2$. Thus the quasi-metric prediction of
${\dot n}_{\rm E}$ is not in conflict with the LLR data if the value of 
${\dot n}_{\rm pE}$ is much smaller than that inferred from using the ELP theory.

It is not clear if the small value $-0.040''$/cy$^2$ of ${\dot n}_{\rm E}$ as
inferred from the ELP theory can be attributed entirely to an expansion of the
Earth's mean orbital radius, or if some part of said value is due to external 
orbit perturbations. We shall assume the former to estimate the expansion 
${\dot a}_{\rm E}$ of the Earth's orbit radius corresponding to the ELP value of
${\dot n}_{\rm E}$. We then compare this estimate to an independent result.
We do the estimate by comparing the ELP and the quasi-metric models using an 
equation similar to equation (78). We also assume that external perturbations 
of the Earth's orbit can be neglected in the quasi-metric model. From Kepler's 
third law, we find an equation similar to equation (78) relating the expansion 
of the Earth's orbit radius ${\dot a}_{\rm E}$ and ${\dot n}_{\rm E}$ as 
calculated from ELP theory to their counterparts as calculated from 
quasi-metric theory. Using Hubble's law to calculate ${\dot a}_{\rm E}$ we find 
a value of about $1.2{\times}10^3$ m/cy from quasi-metric theory. Then, using 
the values for ${\dot n}_{\rm E}$ mentioned above, we estimate ${\dot a}_{\rm E}$ 
as calculated from ELP theory to be about $32$ m/cy from the relationship 
between said quantities. Interestingly, an analysis of all available 
radiometric measurements of distances between the Earth and the major planets,
where radiometric data are compared to calculated distances using planetary 
ephemerides and standard theory, yields a value of $15{\pm}4$ m/cy for 
${\dot a}_{\rm E}$ [16], i.e., about half of the value estimated above. So, 
contrary to what is asserted in [16], an explanation of this result based on 
the cosmic expansion is not at all shown to be inadequate since most, if not 
all, of the substantial difference between ${\dot a}_{\text E}$ as calculated 
from Hubble's law on the one hand and that inferred from radiometric data on 
the other hand, could be due to gross modelling errors. Further evidence for 
the existence of modelling errors due to local cosmic expansion comes from 
optical observations of the Sun, indicating an inconsistency in modern 
ephemerides which may be interpreted as an error of about $1''$/cy in 
$n_{\rm E}$; see, e.g., references [17, 18].

In this section, we have seen that the predicted effects of the cosmic 
expansion on the Earth-Moon and solar system gravitational fields have a number
of observable consequences, none of which is shown to be in conflict with 
observations so far, even though superficially, it would seem that some are.
That is, in every case where there is an apparent conflict between quasi-metric
predictions and observations, the discrepancies can be explained in terms of 
model-dependent assumptions made when analysing the data. In the next section
we will see that a similar situation exists for the predicted versus the
observationally inferred time variation of the gravitational ``constant''.
\subsection{The secular increase of active mass} 
In quasi-metric relativity, active mass varies throughout space-time (but 
{\em not} in the Newtonian limit of the QMF, since this variation is defined 
in terms of a varying scale factor). In particular, for material particles,
there is a secular increase linear in $t$ as seen from equation (8). 
This is equivalent to a secular increase of the gravitational ``constant'' 
$G^{\rm S}_t$. On the other hand, the secular increase of active electromagnetic 
field energy has an extra factor ${\frac{t}{t_0}}$ corresponding to a secular 
increase going as $t^2$ for the second gravitational ``constant'' $G^{\rm B}_t$. 
This means that $G^{\rm B}_t$ and $G^{\rm S}_t$ will be equal for some particular 
cosmic epoch, but the possibility that this is close to the present epoch is 
very unlikely. Therefore, $G^{\rm B}_t$ and $G^{\rm S}_t$ are probably very 
different today. With two different coupling parameters, what is measured in a 
local gravitational test experiment where the gravitational sources do not
follow the cosmic expansion (e.g., a Cavendish experiment), will depend on 
source composition. That is, although electromagnetic field energy does not 
contribute much to source mass, one may in priciple measure $G^{\rm B}_{t_0}$ and 
$G^{\rm S}_{t_0}$ at the present epoch by varying source composition. For the
rest of this paper we will assume that electromagnetic field energy 
contribution to source masses is negligible so that we can set 
$M^{\rm (EM)}_t{\approx}0$. With this approximation, we get the predicted time 
variation of $G^{\rm S}_t$ from equation (8), i.e.,
\eqa
{\frac{G^{\rm S}_{t},_t}{G^{\rm S}_{t}}}={\frac{1}{t}}=(1+O(2))H
{\approx}8{\times}10^{-11}{\,}{\rm yr}^{-1},
\ena
for the present epoch. However, laboratory gravitational experiments are 
nowhere near the experimental accuracy needed to test this prediction. On the 
other hand, space experiments in the solar system (e.g., ranging measurements) 
and observational constraints on solar models from helioseismology are claimed 
to rule out any possible fractional time variation of $G^{\rm S}_t$ larger than 
about $10^{-12}$ yr$^{-1}$. See, e.g., references [19-21] and references therein. 
It thus may appear as the prediction (79) is in conflict with experiment. But 
as we shall see in what follows, this is not the case.

To illustrate the difference between metric and quasi-metric theory when it 
comes to the effects of a varying $G^{\rm S}_t$ on the equations of motion, we
note that in the weak field limit of metric theory we may set $G^{\rm S}_t=
G^{\rm S}_{t_0}+{\dot G}^{\rm S}_{t_0}(t-t_0)+{\cdots}$ directly into the Newtonian 
equation of motion (with $G^{\rm S}_{t_0}=G_{\rm N}$). For an inertial test 
particle this yields (using a Cartesian coordinate system)
\eqa
{\frac{d^2x^j}{dt^2}}=U{\Big [}t,x^k,G^{\rm S}_{t_0}(1+
{\frac{{\dot G}^{\rm S}_{t_0}}{G^{\rm S}_{t_0}}}(t-t_0)+{\cdots}){\Big ]},_j,
\ena
leading to an extra, time-dependent term in the coordinate acceleration of
objects. It is the presence of such an extra term which is ruled out to a
high degree of accuracy according to the space experiments testing the 
temporal variation of $G^{\rm S}_t$. That is, one tests a combination of the 
predicted changes of the solar system scale and orbit periods $T$ which are 
predicted to vary as ${\frac{{\dot T}}{T}}=
-2{\frac{{\dot G}^{\rm S}_t}{G^{\rm S}_t}}$. This follows from Kepler's third law 
since one requires that the conservation of angular momentum takes the form 
${\dot n}=-2{\frac{\dot a}{a}}n$ for any object with mean heliocentric angular 
velocity $n$, mean angular acceleration ${\dot n}$ and fractional change of 
orbit radius ${\frac{\dot a}{a}}$. But this requirement is inconsistent with 
quasi-metric gravity, since we see from equation (35) that the conserved 
quantity is given by ${\frac{t_0}{t}}{\ell}^2n$ (where 
${\ell}{\equiv}{\frac{t}{t_0}}r$ and where corrections of post-Newtonian order 
have been neglected), and not by $r^2n$. This yields 
${\dot n}=-{\frac{\dot a}{a}}n$ and thus equation (79) when applying Kepler's 
third law.

Contrary to metric theory, no such extra term as shown in equation (80) is 
present in the weak field limit of quasi-metric theory since $U$ does not 
depend on $t$. An example of this can be seen from equations (64) and (67), 
where $U{\approx}{\frac{M^{\rm (MA)}_{t_0}G^{\rm S}_{t_0}}{r}}=
{\frac{M_t^{\rm (MA)}G^{\rm S}_{t_0}}{{\ell}}}$ does not depend on $t$. (On the 
other hand, $U$ may depend on $x^0$, but any variation of 
$M^{\rm (MA)}_tG^{\rm S}_{t_0}$ with $x^0$ is not (directly) due to cosmology.) 
However, from equations (70) and (79) we see that in quasi-metric theory, we 
have ${\frac{{\dot T}}{T}}={\frac{{\dot G_{t}}^{\rm S}}{G_{t}^{\rm S}}}$. But as we
have seen in section 4.2, in combination with the predicted scale changes due 
to the cosmic expansion, this is not inconsistent with observations. 

In the weak-field limit of metric theory, one may calculate the effects on
stellar structure coming from a possible variation of $G^{\rm S}_t$. Such effects
are found by putting a variable $G^{\rm S}_t$ directly into the Poisson equation.
That is, a change in $G^{\rm S}_t$ directly induces a change in the Newtonian 
potential yielding a change in star luminosity. Such calculated changes in 
luminosity are tightly constrained from their effects on star models, which can
be compared to observations, e.g., data obtained from helioseismology. On the
other hand, in quasi-metric gravity the effect of the secular increase of 
active mass cannot be separated from the cosmic expansion, so their total 
effect is to {\em decrease} the density (of a body made of ideal gas) with 
cosmic epoch but such that the Newtonian potential is unchanged. (To see how 
this works, recall how equation (12) reduces to equation (47) and take its 
Newtonian limit. Discover that the resulting Poisson equation is unaffected by 
the combination of expansion and increasing $G^{\rm S}_t$.) Thus for a 
main-sequence star there will be approximately no change in luminosity except 
for that due to the increase of scale (i.e., the increase in luminosity due to 
the increasing surface area of the expanding star).

We conclude that all space experimental tests of the secular variation of 
$G^{\rm S}_t$ are based on the assumption that this variation is present 
explicitly in the Newtonian potential. (This underlying assumption is also made
when analysing tests based on stellar structure and in particular restrictions 
coming from helioseismology.) However, said assumption (and in particular 
equation (80)) does not hold in quasi-metric theory. Hence, the interpretations
of these tests are explicitly theory-dependent and the prediction made in 
equation (79) has not been shown to be in conflict with current experimental 
results, despite the variety of tests apparently showing otherwise. Finally, 
note that any cosmological constraints on the secular variation of 
$G^{\rm S}_t$ found within the metric framework are utterly irrelevant for 
quasi-metric theory.
\section{Conclusion}
In this paper, we have shown that according to the QMF, average distances 
within gravitationally bound, metrically static systems are predicted to expand 
according to the Hubble law. Interior to sources the metrically static 
condition applies whenever the equation of state is of the form 
$p{\propto}{\varrho}_{\rm m}$ (fulfilled, e.g., for an ideal gas). When it is 
not, the global cosmic expansion is predicted to induce instabilities 
violating hydrostatic equilibrium. For any such source mass, currents are set 
up to compensate and the system cannot be metrically static. (An example of 
this is a body made of degenerate star matter, e.g. a white dwarf, which is 
predicted to {\em shrink} with epoch.)

According to the QMF, the predicted effects on gravitationally bound systems 
of the global cosmic expansion have a number of observable consequences, none 
of which has been shown to be in conflict with observations. That is, it seems
that at this time no model-independent evidence exists that may rule out the 
possibility that the size of the solar system (measured in atomic units) 
expands according to the Hubble law; on the contrary the quasi-metric model 
fits some observational data more naturally than traditional models do.
(But note that predictions coming from the QMF fit these data naturally only 
as long as the data are analysed and interpreted in a manner consistent with 
the QMF.)

Some examples of observations being naturally explained within the non-metric 
sector of the QMF have been discussed in this paper; e.g., the spin-down of 
the Earth [10, 11], the recession of the Moon and its mean acceleration 
[13, 14], and the newly discovered secular increase of the astronomical unit 
[16]. Also the so-called ``Pioneer effect'' has a natural explanation within 
the QMF [22]. Thus the non-metric sector of the QMF has considerable 
predictive power in the solar system, since it makes it possible to explain 
from first principles a number of seemingly unrelated phenomena as different 
aspects of the same model. On the other hand, explanations of these phenomena 
coming from standard theory are invariably {\em ad hoc}; such explanations 
always involve free parameters and mechanisms invented to explain each 
phenomenon separately. Such an approach is untenable according to Occam's 
razor. 

So, fact is that several observations in the solar system represent evidence 
that space-time is quasi-metric. Moreover, metric gravity (and General 
Relativity in particular) fails to address the challenge represented by these 
observations. And the main reason that this challenge has not been recognized 
as important, is that experimental gravity in the solar system is analysed 
within a weak-field formalism (the so-called PPN-formalism) where it is 
inherently assumed that space-time must be modelled as a pseudo-Riemannian 
manifold, and consequently that the cosmic expansion should be unmeasurably 
small at the scale of the solar system. This situation may change in the 
future, when solar system gravitational experiments reach a precision level 
where the quasi-metric effects can no longer reasonably be ``explained'' by 
adding {\em ad hoc} hypotheses to metric theory.
\\ [4mm]
{\bf Acknowledgment} \\ [1mm]
I wish to thank Dr. K{\aa}re Olaussen for making a critical review
of the manuscript. \\ [4mm]
{\bf References} \\ [1mm]
{\bf [1]} F.I. Cooperstock, V. Faraoni, D.N. Vollick, {\em ApJ} {\bf 503}, 61
(1998) (astro-ph/9803097). \\
{\bf [2]} T. Ekholm, Yu. Baryshev, P. Teerikorpi, M.O. Hanski, G. Paturel, \\
\hspace*{6.3mm} {\em A{\&}A} {\bf 368} L17 (2001) (astro-ph/0103090). \\
{\bf [3]} D. {\O}stvang, {\em Grav. \& Cosmol.} {\bf 11}, 205 (2005)
(gr-qc/0112025). \\
{\bf [4]} D. {\O}stvang, {\em Doctoral Thesis}, NTNU (2001) (gr-qc/0111110).\\
{\bf [5]} D. {\O}stvang, D. {\O}stvang, {\em Grav. \& Cosmol.} {\bf 12}, 262 
(2006) (gr-qc/0303107). \\
{\bf [6]} P. Wesson, {\em Cosmology and Geophysics}, Adam Hilger Ltd, (1978).\\
{\bf [7]} S. Weinberg, {\em Gravitation and Cosmology}, John Wiley ${\&}$ Sons,
Inc. (1972). \\
{\bf [8]} P.D. Mannheim, {\em Found. Phys.} {\bf 24}, 487 (1994). \\
{\bf [9]} D. {\O}stvang, {\em Acta Physica Polonica B} {\bf 39}, 1849 (2008)
(gr-qc/0510085). \\
{\bf [10]} F.R. Stephenson and L.V. Morrison, {\em Phil. Trans. R. Soc. Lond.}
{\bf A313}, 47 (1984). \\
{\bf [11]} F.R. Stephenson and L.V. Morrison, {\em Phil. Trans. R. Soc. Lond.}
{\bf A351}, 165 (1995). \\
{\bf [12]} G.E. Williams, {\em Geophysical Research Letters} {\bf 24}, 421
(1997). \\
{\bf [13]} J. Chapront, M. Chapront-Touz\'{e}, G. Francou, {\em  A{\&}A} 
{\bf 387}, 700 (2002). \\
{\bf [14]} J.O. Dickey {\em et al.}, {\em Science} {\bf 265}, 482 (1994). \\
{\bf [15]} W.H. Munk, {\em Prog. Oceanog.} {\bf 40}, 7 (1997). \\
{\bf [16]} G.A. Krasinsky and V.A. Brumberg, 
{\em Celes. Mech. {\&} Dyn. Astron.} {\bf 90}, 267 (2004). \\
{\bf [17]} P. Stumpff and J.H. Lieske, {\em  A{\&}A} {\bf 130}, 211 (1984). \\
{\bf [18]} P.K. Seidelmann, in {\em IAU Symposium No. 152}, 49 (1992). \\
{\bf [19]} J-P. Uzan, {\em Rev. Mod. Phys.} {\bf 75}, 403 (2003) 
(hep-ph/0205340). \\
{\bf [20]} T. Chiba, gr-qc/0110118. \\
{\bf [21]} J.G. Williams, S.G. Turyshev, D.H. Boggs,
{\em Phys. Rev. Lett.} {\bf 93}, 261101 (2004). \\
{\bf [22]} D. {\O}stvang, {\em Class. Quantum Grav.} {\bf 19}, 4131 (2002)
(gr-qc/9910054).
\end{document}